\documentclass[11pt]{article}
\usepackage[utf8]{inputenc}
\usepackage[left=2.5cm,right=2.5cm,top=2.5cm,bottom=2.5cm]{geometry}
\usepackage{lmodern}

\usepackage{setspace} 
\usepackage{hhline} 
\usepackage{float} 
\usepackage{array} 
\usepackage{eurosym} 
\usepackage{arydshln} 
\usepackage{longtable} 
\usepackage{booktabs}
\usepackage{csquotes} 
\usepackage{nccmath} 
\usepackage{booktabs} 
\usepackage{xcolor} 
\usepackage{multirow} 
\usepackage{bm} 
\usepackage{physics} 
\usepackage{systeme} 
\usepackage{mathtools} 
\usepackage{amsmath, amsthm, amssymb, amsfonts} 
\usepackage{graphicx} 
\usepackage{pgfplots} 
\pgfplotsset{compat=1.17} 
\usepackage{tikz} 
\usetikzlibrary{intersections,calc} 
\usepackage[labelfont=bf,textfont=it]{caption} 
\usepackage{soul} 
\usepackage{sidecap} 
\usepackage[space]{grffile} 
\usepackage{subcaption} 
\usepackage{makebox} 
\usepackage{abraces} 
\usepackage{enumitem} 
\usepackage{comment} 
\usepackage{cancel} 
\usepackage{accents} 
\usepackage{stackrel} 
\usepackage[export]{adjustbox}
\usepackage{centernot} 
\usepackage{scalerel} 
\usepackage{makeidx}
\usepackage{hhline}
\usepackage{chngcntr}
\usepackage[flushleft]{threeparttable}
\usepackage{listings}
\usepackage[round]{natbib}
\usepackage{hyperref,etoolbox}
\hypersetup{
    colorlinks=true,
    linkcolor=black,
    filecolor=magenta,
    citecolor=black, 
    urlcolor=black,
    }
\makeatletter
\def\@footnotecolor{red}
\define@key{Hyp}{footnotecolor}{%
 \HyColor@HyperrefColor{#1}\@footnotecolor%
}
\patchcmd{\@footnotemark}{\hyper@linkstart{link}}{\hyper@linkstart{footnote}}{}{}
\makeatother
\hypersetup{footnotecolor=black}
\usepackage{pdflscape}

\usepackage{chngcntr}
\usepackage[ruled,vlined]{algorithm2e}
\SetKwInput{KwInput}{Input}                
\SetKwInput{KwOutput}{Output}

\newlength{\bibitemsep}
\newlength{\bibparskip}
\let\oldthebibliography\thebibliography
\renewcommand\thebibliography[1]{%
  \oldthebibliography{#1}%
  \setlength{\parskip}{\bibitemsep}%
  \setlength{\itemsep}{\bibparskip}%
}

\usetikzlibrary{arrows,decorations.markings}
\usetikzlibrary{decorations.pathreplacing}
\tikzset{myarr/.style={decoration={markings, mark=between positions 0 and 1 step 4mm with {\arrow{stealth}},},postaction=decorate}}
\tikzset{myarrrev/.style={decoration={markings, mark=between positions 0 and 1 step 4mm with {\arrowreversed{stealth}},},postaction=decorate}}
\DeclareCaptionLabelFormat{subfig}{\figurename #1~\thefigure.\alph{subfigure}:}

\newcommand{\parT}[1]{\left(#1\right)}
\newcommand{\parQ}[1]{\left[#1\right]}

\newcommand{\ubar}[1]{\underaccent{\bar}{#1}}

\newcommand{\I}{\mathbb{I}}

\newcommand*{\giv}{\hspace{1pt}|\hspace{1pt}}

\newcommand{\bs}[1]{\boldsymbol{#1}}

\newcommand{\STAB}[1]{\begin{tabular}{@{}c@{}}#1\end{tabular}}

\theoremstyle{definition}

\newtheorem*{assumption*}{\assumptionnumber}
\providecommand{\assumptionnumber}{}


\theoremstyle{definition}

\theoremstyle{definition}

\theoremstyle{definition}

\theoremstyle{definition}

\theoremstyle{definition}

\theoremstyle{definition}

\definecolor{blue}{RGB}{0,114,178}
\definecolor{red}{RGB}{204,51,17}
\definecolor{yellow}{RGB}{240,228,66}
\definecolor{green}{RGB}{0,158,115}

\title{Heterogeneous economic growth vulnerability across Euro Area countries under stressed scenarios}
\author{\normalsize Claudio Lissona\footnote{Corresponding author, e-mail: \url{claudio.lissona2@unibo.it}} \footnote{Department of Economics, University of Bologna}
\and
\normalsize Esther Ruiz\footnote{Department of Statistics, Universidad Carlos III de Madrid}
}

\date{\normalsize\today}
\begin{document}

\maketitle

\begin{abstract}
\noindent  We analyse economic growth vulnerability of the four largest Euro Area (EA) countries under stressed macroeconomic and financial conditions. Vulnerability, measured as a lower quantile of the growth distribution conditional on EA-wide and country-specific underlying factors, is found to be higher in Germany, which is more exposed to EA-wide economic conditions, and in Spain, which has large country-specific sectoral dynamics. We show that, under stress, financial factors amplify adverse macroeconomic conditions. Furthermore, even severe sectoral (financial or macro) shocks, whether common or country-specific, fail to fully explain the vulnerability observed under overall stress. Our results underscore the importance of monitoring both local and EA-wide macro-financial conditions to design effective policies for mitigating growth vulnerability.

\medskip\noindent
{\it Keywords}: Economic scenarios, GaR, GiS, Multi-level dynamic factor model, EM algorithm, Quantile regression.\\
\end{abstract}

\vspace{10pt}

\newpage

\section{Introduction}

Understanding the risks of severe economic downturns--referred to as downside risks to economic growth or growth vulnerability--has become a central concern for policymakers and market participants \citep[see, e.g.,][]{kilian2008central,alessi2014central}. The potential for sudden, large and unexpected contractions in economic activity poses significant challenges for macro-financial stability. As a consequence of the influential work of \citet*{adrian2019vulnerable}, it is popular to quantify downside risks to economic growth with Growth-at-Risk (GaR), which nowadays is part of the toolbox of academics and policy makers  \citep{prasad2019growth,suarez2022growth,furno2024nowcasting}.

While a substantial body of research on growth vulnerability focuses on the US, there is growing interest in examining such risks within the Euro Area (EA). Accurately assessing and anticipating growth vulnerability is critical for the timely implementation of both EA-wide and country-specific policies aimed at mitigating the impact of adverse shocks. Given the interconnectedness of EA economies together with the diverse structural conditions across member countries, it is essential to develop models that can capture both global (EA-wide) and country-specific factors influencing growth vulnerability within EA countries.



In this paper, we analyse economic growth vulnerability across the four largest EA economies--Germany (DE), France (FR), Italy (IT), and Spain (ES).\footnote{As of 2023, Germany, Spain, France, and Italy account for approximately two-third of overall GDP of the EA and are frequently used as representative countries for economic analysis; see, for example, \citet{angelini2019mind}.} For each country, we model the distribution of GDP growth conditional on both global EA and country-specific macro and financial conditions and compute the corresponding GaRs. By doing so, we are able to evaluate the heterogeneous effects of global and country-specific macro and financial conditions on vulnerable growth across the largest economies in the EA. We then generate scenarios for growth vulnerability when these conditions are under stress to understand how downside risks to GDP growth differ from those observed during \enquote{normal} times. 

Our methodology relies on three main steps. First, for each country, we compute the GaR following \citet*{adrian2019vulnerable}, estimating the growth density after fitting factor-augmented quantile regressions (FA-QRs) separately for each country. The factors used in the FA-QRs, which quantify EA-wide and country-specific macroeconomic and financial conditions, are extracted using a multilevel dynamic factor model (ML-DFM) estimated via the EM algorithm \citep{barigozzi2022quasi,delle2022common}. Specifically, we extract pervasive EA-wide global factors and semi-pervasive country-specific factors from a large dataset of macroeconomic and financial indicators, as provided by \citet*{barigozzi2024large}. This approach allows us to disentangle common dynamics, shared across EA economies, from country-specific macro-financial conditions. We document the presence of macroeconomic and financial factors, which capture comovements across EA countries, serving as proxies for aggregate macro-financial conditions in the EA. Additionally, the inclusion of country-specific factors provides a detailed view of local sectoral dynamics, allowing for a comprehensive analysis of how EA-wide and local macro-financial conditions jointly shape economic growth vulnerability in each country.

Second, \citet*{gonzalez2024expecting} show that, under stressed conditions, which are critical for policymakers, standard growth vulnerability measures like GaR tend to underestimate downside risks to growth. In the context of Principal Components (PC) factor extraction, they introduce Growth-in-Stress (GiS) as a complementary measure that assesses vulnerability when the underlying factors are jointly stressed using their probability countours. In this paper, we extend the methodology proposed by \cite{gonzalez2024expecting} to factors extracted from ML-DFMs using the EM algorithm instead of PC. For each of the four EA economies considered, we construct scenarios to evaluate growth vulnerability under economic stress, generating scenarios for extreme deviations in macroeconomic and financial conditions, both within and across countries and obtain the GiS under these scenarios. Our findings reveal significant heterogeneity across countries, with those more exposed to EA-wide conditions, such as Germany, and those driven by strong country-specific dynamics, such as Spain, exhibiting higher levels of growth vulnerability. This underscores the importance of considering both common and local dynamics in evaluating risks to domestic economic growth.

Moreover, we make a further step to disentangle the drivers of growth vulnerability within each country by exploring alternative stress-testing scenarios. We move beyond joint stress scenarios by separately stressing alternative factors. First, we consider a scenario where only macroeconomic factors are stressed, while financial factors remain at their average levels. This exercise reveals that vulnerability is systematically underestimated when financial conditions are not included in the stress test, highlighting the crucial amplifying role that financial stress plays in driving adverse economic outcomes. Second, we examine the impact of large sectoral shocks, both common and country-specific, to determine whether extreme shocks to individual sector can replicate the large vulnerabilities observed when the economy is jointly stressed. This analysis provides novel evidence on the amplifying role of financial variables in adverse economic conditions and underscores the risks to growth posed by economy-wide stress compared to local, yet very large, sector-specific shocks. Our findings show that large sectoral shocks fail to produce the same levels of vulnerability as those observed under joint stress, reinforcing the idea that growth vulnerability across EA countries is largely shaped by the interaction between both EA-wide and country-specific macroeconomic, and financial conditions.

Our paper is related with previous work on the quantification of growth vulnerability across different dimensions. One of the earliest attempts to measure EA growth vulnerability is \citet{proietti2017euromind}, who construct a monthly density indicator for GDP. Later, \citet{figueres2020vulnerable} and \citet{ferrara2022high} employed the GaR framework to model tail risk in the EA, documenting the importance of financial indicators to model downturn risk for GDP growth.\footnote{Specifically, they consider the Composite Indicator of Systemic Stress (CISS), obtained by non-linear aggregation of individual financial indicators; see \citet{hollo2012ciss} for a description of CISS.} Similarly, \citet{szendrei2023revisiting} employ a frequentist Adaptive LASSO approach to identify key variables driving different parts of the EA growth distribution, emphasizing the significance of financial, particularly bank-related, variables. Conversely, \citet{lhuissier2022financial} proposes a regime-switching skew-normal model for the EA growth distribution. Unlike other studies, his findings indicate a limited contribution of financial variables, when included in time-varying probabilities used to anticipate shifts in growth skewness. Overall, a broad literature underscores the joint role of macroeconomic and financial conditions in shaping macroeconomic tail risks, extending beyond economic growth alone. For instance, \citet{botelho2024labour} examine tail risk in the EA (and US) labour market, modelling it as a function of both real and financial indicators and emphasizing the role of non-linearities in the transmission of shocks to unemployment. Additionally, \citet{morana2024new} propose constructing composite indicators for the EA that integrate macroeconomic and financial drivers, capturing both short- and long-term fluctuations in these variables.

While substantial research has focused on modelling growth vulnerability for the EA as a whole, EA countries exhibit significant heterogeneity in labour markets, financial structures, and fiscal positions \citep{gilchrist2018credit,hoffmann2020financial}. Within EA members states, economic vulnerabilities may differ substantially due to various factors, which may reflect either EA-wide or country-specific macro-financial conditions, leading to heterogeneous effects on downside growth risks across countries; see, for example, \citet{eickmeier2009comovements}. The COVID-19 shock, for instance, despite being a common shock, had asymmetric effects on EA countries in both timing and magnitude—an aspect that would be averaged out when focusing solely on aggregate EA growth. Yet, the analysis of growth vulnerability at the individual country level remains limited.

Our work contributes to this literature along different dimensions. First, we move beyond a standard aggregate approach by modelling economic growth vulnerability within each of the main EA economies as a function of both common and country-specific economic conditions. By doing so, we are able to account for potential cross-country spillovers, as well as local dynamics which would be otherwise neglected  in an EA-wide analysis. Second, we explicitly model economic growth vulnerability as a function of macroeconomic and financial conditions, which allows us to assess the relative contribution of each sector to growth risks. We achieve this by leveraging multiple economic sources within our ML-DFM, effectively incorporating all relevant information across sectors and countries. Finally, our stressed scenarios provide valuable insights that extend beyond standard measures of downside GDP risk. Specifically, they highlight the role of financial variables in amplifying economic downturns and the significance of local dynamics in shaping the exposure of each country to macro-financial risks. To the best of our knowledge, we are the first to model downside risks to GDP at the individual EA country level while simultaneously accounting for both EA-wide and country-specific conditions. As far as we are aware, only \citet{busetti2021time} explicitly model the growth distribution for Italy, using expectile regressions based on both domestic and international real and financial indicators. Furthermore, in contrast to the existing literature, we do so by utilizing a novel high-dimensional dataset.

In light of the literature, the framework adopted in this paper offers a comprehensive approach to understanding growth risks across EA countries, highlighting the need to monitor both EA-wide and local macro-financial conditions. Our results emphasize the systemic nature of growth risks, providing evidence on the pivotal role that the interplay between macroeconomic and financial conditions plays in shaping vulnerabilities across economies. We offer valuable insights for policymakers, emphasizing the importance of coordinated macro-prudential policies, alongside monetary and fiscal tools, tailored to address both common and country-specific vulnerabilities. Such measures are essential to mitigate growth risks effectively and ensure economic stability within the EA.

The rest of the paper is structured as follows. In Section \ref{sec::method} we describe the methodology and estimation strategy. Section \ref{sec::DA} describes the data used in the analysis. The main results of the paper appear in Section \ref{sec::res}, which deals with the empirical estimation of growth vulnerability in each of the four largest EA economies and presents different stressed scenarios. Section \ref{sec::conclude} concludes.

\section{Methodology}
\label{sec::method}

In this section, we describe the methodology used to estimate the conditional distribution of GDP growth in each country, and the corresponding GaR and GiS measures of vulnerability. On top of describing the FA-QR model, we also describe the ML-DFM used to extract the latent factors driving macroeconomic and financial conditions.

\subsection{The conditional distribution of GDP growth: GaR and GiS}
\label{sbsec::condGDP}

We characterize the distribution of GDP growth conditional on underlying macroeconomic and financial factors under two distinct scenarios: (i) during \enquote{normal} times, with the latent factors held at their average levels, and (ii) under stress, when the economic conditions represented by the factors are subject to significant macroeconomic and/or financial pressures.

\subsubsection{Conditional distribution in normal times: growth at risk (GaR)}
\label{sbsbsec::GaR}

To analyse the relationship between GDP growth and past macroeconomic and financial conditions, we fit a FA-QR model for each country, with the quantiles modelled as functions of a few latent macro/financial factors capturing both EA-wide comovements across countries, as well as domestic dynamics within each country. The four EA countries--Germany (DE), Spain (ES), France (FR), and Italy (IT)--are indexed by $c\in \mathcal{C}:=$ \{DE,ES,FR,IT\}. Let $y_{t}^{(c)}$ denote the quarterly GDP growth rate in period $t$ for country $c\in\mathcal{C}$, and let $\mathbf{F}_t^{(c)} = (\mathbf{f}_t^{\text{(F)}'},\mathbf{f}_t^{\text{(M)}'}, \mathbf{f}_{t}^{(c,\text{F})'},\mathbf{f}_{t}^{(c,\text{M})'})'$ denote the vector of factors assumed to characterize its distribution. The $r^{(\text{F})}$ factors in $\mathbf{f}_t^{(\text{F})}$ are financial factors common to all four countries, while the $r^{(c,\text{F})}$ factors in $\mathbf{f}_t^{(c,\text{F})}$ are financial factors specific to country $c$. Similarly, $\mathbf{f}_t^{(\text{M})}$ and $\mathbf{f}_t^{(c,\text{M})}$, with dimensions $r^{(\text{M})}$ and $r^{(c,\text{M})}$, respectively, are common and country-specific macroeconomic factors, respectively. The total number of factors is $r = r^{(\text{F})} + \sum_{c\in\mathcal{C}}r^{(c,\text{F})} + r^{(\text{M})} + \sum_{c\in \mathcal{C}}r^{(c,\text{M})}$. Let the two economic sectors--macroeconomic (M) and financial (F)--be indexed by $s\in\mathcal{S}:=$ \{F, M\}. For each country within $\mathcal{C}$, and quantile $\tau = 0.05,0.10,\ldots,0.95$, we fit the following FA-QR model:\footnote{Although the analysis in this paper focus on one-step-ahead growth, it can be easily extended to other
multi-period forecast horizons.}


\begin{equation}
\label{eq::Qreg}
q_{\tau}(y_{t+1}^{(c)}, \giv y_t^{(c)}, \mathbf{F}_t^{(c)})\ =\ \mu^{(c,\tau)} + \phi^{(c,\tau)}y_{t}^{(c)} + \sum_{s\in\mathcal{S}}\bs{\beta^}{(c,s,\tau)\prime}\mathbf{f}_{t}^{(s)} + \sum_{s\in\mathcal{S}}\bs{\gamma}^{(c,s,\tau)\prime}\mathbf{f}_{t}^{(c,s)} 
\end{equation}
where $q_{\tau}(y_{t+1}^{(c)}, \giv y_t^{(c)},\mathbf{F}_t^{(c)})$ is the one-step-ahead forecast of the $\tau-$quantile of the conditional distribution of GDP growth for country $c$, while $\mu^{(c,\tau)}$, $\phi^{(c,\tau)}$, $\bs{\beta}^{(c,s,\tau)} = \parT{\beta_1^{(c,s,\tau)},\ldots,\beta_{r^{(s)}}^{(c,s,\tau)}}$ and $\bs{\gamma}^{(c,s,\tau)} = \parT{\gamma_1^{(c,s,\tau)},\ldots,\gamma_{r^{(c,s)}}^{(c,s,\tau)}}$ are parameters to be estimated.  Since the factors are unobserved, they are replaced in practice by their estimated counterparts. The model employed to extract the factors and their estimation are treated in detail in Section \ref{sbsec::MLVDFM}.

The FA-QR model in \eqref{eq::Qreg} is a flexible yet parsimonious tool that effectively captures potential non-linear and asymmetric relationships between economic growth and the underlying macroeconomic and financial conditions, with EA-wide economic conditions represented by the common factors and country-specific dynamics captured by the country-level factors. This dual perspective is crucial for evaluating the presence, if any, of heterogeneous responses to common (EA-wide) shocks and their interactions with local macroeconomic and financial conditions. 

The parameters of model \eqref{eq::Qreg} are estimated following the methodology introduced in \citet{koenker1978regression}, using the algorithm of \citet{koenker1987algorithm}, with the corresponding standard errors computed as described in \citet{koenker2005quantile} assuming i.i.d. errors; see \citet{bai2008extremum} for the consistency of the estimated parameters, and \citet{ando2011quantile}, and \citet{giglio2016systemic} for the consistency of the forecasts of the quantiles when the factors are extracted using PC. The goodness of fit of the model is assessed using the $R^1$ measure described in \cite{koenker1999goodness}. Specifically, let \begin{footnotesize}$\hat{v}_{t+1}^{(c,\tau)} = y_{t+1}^{(c)} - \widehat{\mu}^{(c,\tau)} - \widehat{\phi}^{(c,\tau)}y_{t}^{(c)} - \sum_{s\in\mathcal{S}}\widehat{\bs{\beta}}^{(c,s,\tau)}\mathbf{f}_{t}^{(s)} - \sum_{s\in\mathcal{S}}\widehat{\bs{\gamma}}^{(c,s,\tau)}\mathbf{f}_{t}^{(c,s)}$\end{footnotesize} denote the residuals from model (1) fitted to the $\tau$-quantile of growth in country $c$. For each quantile and country, we compute \begin{footnotesize}$R^1(c,\tau) = 1-\frac{\sum_{t=2}^{T-h} \hat{v}_t^{(c,\tau)}[\tau \I(\hat{v}_t^{(c,\tau)} \geq 0)+(\tau-1) \I(\hat{v}_t^{(c,\tau)}<0)]}{\sum_{t=2}^{T-h} (y_{t}^{(c)}-\bar{y}^{(c)})[\tau(\I(y_{t}^{(c)} \geq \bar{y}^{(c)})+(\tau-1) \I(y_{t}^{(c)}<\bar{y}^{(c)})]}$\end{footnotesize}, where $\bar{y}^{(c)}$ is the sample mean of $y_t$ for country $c$ and $\I(\cdot)$ is the indicator function, which takes a value of 1 if its argument is true, and 0 otherwise. 

After estimating model (1), we follow \citet{adrian2019vulnerable} and obtain a smooth estimate of the conditional distribution of growth for each country and period $t$ by fitting the Skewed-t distribution introduced by \citet{azzalini1985class} to the estimated $\tau = 0.25,0.5,0.75,0.95$ quantiles\footnote{The results are nearly identical to those obtained by using additional quantiles as, for example, $\tau = 0.05,0.10,0.90$.}. The estimated conditional distribution of GDP growth for each country is denoted as $\widehat{f}_0(y_{t+1}^{(c)}\giv, y_{t}^{(c)}, \mathbf{F}_t^{(c)})$. Finally, we compute the one-step-ahead $\text{GaR}_t$ as the $5\%$ quantile of this distribution.

\subsubsection{Conditional distribution under stress: growth in stress (GiS)}
\label{sbsbsec::GiS}

In contrast to GaR, which assesses growth vulnerability under \enquote{normal} conditions--i.e. when the factors are fixed at their average values--GiS evaluates growth vulnerability under stressed conditions by considering factors that deviate significantly from their average value; see \cite{gonzalez2019growth} and \cite{gonzalez2024expecting}. Indeed, extreme structural economic conditions tend to exacerbate growth vulnerability, a phenomenon that is likely underestimated by GaR.

In order to estimate the density of growth under stressed factors, we focus on the 5\% quantile and find the value of the factors, when constrained at their $\alpha$-probability contour, that minimize  $q_{0.05}(y_{t+1}^{(c)}\giv y_{t}^{(c)}, \mathbf{F}_t^{(c)})$ as follows:\\
\begin{equation}
\label{eq::conmin}
\min_{\mathbf{F}_t^{(c)}}\quad q_{0.05}(y_{t+1}^{(c)} \giv y_{t}^{(c)}, \mathbf{F}_t^{(c)}) \hspace{25pt} \text{s.t.} \quad g(\mathbf{F}_t^{(c)},\alpha) = 0
\end{equation}\\
where $g(\mathbf{F}_t^{(c)},\alpha)$ represents the $\alpha$-contour of the factors for each country, with $\alpha$ being a predetermined probability level that defines the severity of the underlying economic stress. Higher values of $\alpha$ correspond to more extreme underlying stress levels for the macroeconomic and financial conditions represented by factors. In this paper, we consider $\alpha = 0.95$. Furthermore, note that the joint probability contours $g(\cdot)$ for each country are shaped by
both EA-wide and country-specific factors. Therefore, for the same level of underlying stress, the composition of underlying economic factors can be different for each country. These differences depend on the transmission of common shocks to individual countries and the significance of local macro-financial conditions within each economy.

In practice, when the number of factors is larger than 2, as in our case, the minimization problem in \eqref{eq::conmin} is solved using the binary mesh algorithm proposed by \cite{flood2015systematic}.\footnote{This algorithm is widely employed in stress-testing exercises by financial institutions in order to leverage the dependence structure among the underlying factors to create plausible and severe scenarios with minimal parametrization; see on Basel Committee \citet{basel2009principles}. The binary mesh algorithm requires selecting a tuning parameter $\delta$ that determines the granularity of the mesh. Higher values of $\delta$ increase the number of scenarios considered, at the cost of a greater computational burden. Consistent with \citet{gonzalez2024expecting}, we set $\delta = 8$, corresponding to $\approx$3000 scenarios. Results are unchanged with higher mesh levels (e.g. $\delta = 10$, with $\approx$ 5500 scenarios).} Finally, the factors that solve the minimization problem in \eqref{eq::conmin} are substituted into the estimated FA-QRs for $\tau =$ 0.25,0.50,0.75 and 0,95 to obtain the corresponding quantiles to which we fit the Skew-t distribution. This approach yields a conditional density of GDP growth under stressed conditions. For a given probability level $\alpha$ and each country $c$, we denote this estimated density as $\widehat{f}_{\alpha}(y_{t+h}^{(c)}\giv y_{t}^{(c)}, \mathbf{F}_t^{(c)})$. The one-step-ahead $\text{GiS}_t$ is defined as the 5\% quantile of this density. The GiS provides a measure of the vulnerability of economic growth to adverse structural economic conditions under different scenarios, serving as a valuable tool for policymakers.

\subsection{Factor extraction: Multilevel DFM}
\label{sbsec::MLVDFM}

We fit two separate ML-DFMs to the sets of macroeconomic (M) and financial (F) variables to extract the corresponding latent factors. Within each ML-DFM, we assume that the block structure of the data is known, with the factors being either common to all countries or specific of each country. 
 We test for the presence of common factors overlapping both sets of variables by implementing the bootstrap tests proposed by \citet{gonccalves2025bootstrap}, who after estimating factors separately in each set of variables, test whether the factors in different sets are correlated. Denote as $\mathbf{x}_{t}^{\scaleto{(s)}{6pt}} = (\mathbf{x}_{t}^{\scaleto{(\text{DE},s)}{6pt}'},\mathbf{x}_{t}^{\scaleto{(\text{ES},s)}{6pt}'},\mathbf{x}_{t}^{\scaleto{(\text{FR},i)}{6pt}'},\mathbf{x}_{t}^{\scaleto{(\text{IT},i)}{6pt}'})'$ the $N_s-$dimensional vector of variables for sector $s \in \mathcal{S}$ at time $t$, such that, for country $c\in \mathcal{C}$, $\mathbf{x}_{t}^{\scaleto{(c,s)}{6pt}} = (x_{1,t}^{\scaleto{(c,s)}{6pt}},\ldots,x_{n_{c,s},t}^{\scaleto{(c,s)}{6pt}})'$ represents the $N_{c,s}-$dimensional vector of country-sector specific variables. The model for each sector $s$ can be expressed as follows:\\
\begin{align}
\label{eq::mlDFM}
\mathbf{x}_t^{(s)}\ &=\ 
\parQ{
\begin{array}{l|cccc}
\bs{\Lambda}^{\scaleto{(\text{DE},s)}{6pt}} & \bs{\Psi}^{\scaleto{(\text{DE},s)}{6pt}} & \mathbf{0} & \mathbf{0} & \mathbf{0} \\[3pt]
\bs{\Lambda}^{\scaleto{(\text{ES},s)}{6pt}} & \mathbf{0} & \bs{\Psi}^{\scaleto{(\text{ES},s)}{6pt}} & \mathbf{0} & \mathbf{0} \\[3pt]
\bs{\Lambda}^{\scaleto{(\text{FR},s)}{6pt}} & \mathbf{0} & \mathbf{0} & \bs{\Psi}^{\scaleto{(\text{FR},s)}{6pt}} & \mathbf{0} \\[3pt]
\bs{\Lambda}^{\scaleto{(\text{IT},s)}{6pt}} & \mathbf{0} & \mathbf{0} & \mathbf{0} & \bs{\Psi}^{\scaleto{(\text{IT},s)}{6pt}}
\end{array}
}
\parQ{
\begin{array}{c}
\mathbf{f}_t^{\scaleto{(s)}{6pt}}\\
\hline\\[-12pt]
\mathbf{f}_t^{\scaleto{(\text{DE},s)}{6pt}}\\
\mathbf{f}_t^{\scaleto{(\text{ES},s)}{6pt}}\\
\mathbf{f}_t^{\scaleto{(\text{FR},s)}{6pt}}\\
\mathbf{f}_t^{\scaleto{(\text{IT},s)}{6pt}}\\
\end{array}
}
+
\bs{\xi}_{t}^{(s)}\\[10pt]
\mathbf{f}_t^{(s)}\ &=\ \mathbf{A}^{(s)}(L)\mathbf{f}_{t-1}^{(s)} + \mathbf{u}_t^{(s)}  \nonumber \\[5pt] 
\mathbf{f}_t^{\scaleto{(c,s)}{6pt}}\ &=\ \mathbf{B}^{\scaleto{(c,s)}{6pt}}(L)\mathbf{f}^{\scaleto{(c,s)}{6pt}}_{t-1} + \mathbf{v}_t^{\scaleto{(c,s)}{6pt}}  \nonumber 
\end{align}\\
where the factors are defined as in model \eqref{eq::Qreg} and $\bs{\Lambda}^{(s)} = (\bs{\Lambda}^{\scaleto{(\text{DE},s)}{6pt}},\bs{\Lambda}^{\scaleto{(\text{ES},s)}{6pt}},\bs{\Lambda}^{\scaleto{(\text{FR},s)}{6pt}},\bs{\Lambda}^{\scaleto{(\text{IT},s)}{6pt}})$ is the $N^{(s)} \times r^{(s)}$ matrix of loadings for sector $s$ common to all countries, with $\bs{\Lambda}^{\scaleto{(c,s)}{6pt}} = (\bs{\lambda}_{1}^{\scaleto{(c,s)}{6pt}'},\ldots, \bs{\lambda}_{n_c}^{\scaleto{(c,s)}{6pt}'})'$ being the $N^{(c,s)}\times r^{(c,s)}$ matrix of loadings for country $c$. Similarly, $\mathbf{f}_{t}^{\scaleto{(c,s)}{6pt}}$ is the $r^{(c,s)}-$dimensional vector of loadings corresponding to the  country-specific factors. Finally, $\bs{\xi}_t^{(s)} = (\bs{\xi}_t^{\scaleto{(\text{DE},s)}{6pt}'},\bs{\xi}_t^{\scaleto{(\text{ES},s)}{6pt}'},\bs{\xi}_t^{\scaleto{(\text{FR},s)}{6pt}'}$, $\bs{\xi}_t^{\scaleto{(\text{IT},s)}{6pt}'})'$, is the $N^{(s)}$-dimensional vector of white noise idiosyncratic components, with $\bs{\xi}_t^{\scaleto{(c,s)}{6pt}} = (\xi_{1,t}^{\scaleto{(c,s)}{6pt}},\ldots, \xi_{n_{c,s},t}^{\scaleto{(c,s)}{6pt}})'$; see, among others, \citet{breitung2016analyzing}, \citet{choi2018multilevel}, \citet{andreou2019inference}, \citet{freyaldenhoven2022factor} and \citet{gonccalves2025bootstrap} for the characterization of the ML-DFM. The identification of factor and loadings requires that, within each block of variables, the factors should be orthonormal with the corresponding loadings being orthogonal; see \citet{breitung2016analyzing}.\footnote{Note that  \citet{freyaldenhoven2022factor} also requires orthogonality between blocks.} Furthermore, we assume that the dynamic dependence of the factors is described by a VAR(1) model with $\mathbf{A}^{(s)}$ and $\mathbf{B}^{\scaleto{(c,s)}{6pt}}$ being $r^{(s)} \times r^{(s)}$ and $r^{(c,s)}\times r^{(c,s)}$, respectively. The $r^{(s)}$- and $r^{(c,s)}$-dimensional vectors $\mathbf{u}_t^{(s)}$ and $\mathbf{v}_t^{\scaleto{(c,s)}{6pt}}$ are white noise residuals; see Appendix \ref{app::modest} for the complete state-space representation of the ML-DFM in (\ref{eq::mlDFM}).

The number of factors within each sector and country is determined by examining the corresponding scree plots, as well as the covariance matrix of estimated idiosyncratic components, to ensure that no relevant comovements are left unexplained by the model.\footnote{Several procedures to determine the number of factors in ML-DFMs have been proposed in the literature; see, for example, \citet{freyaldenhoven2022factor} and \citet{choi2023canonical}.}

The ML-DFM in \eqref{eq::mlDFM} is finally estimated using the Expectation-Maximization (EM) algorithm, as outlined by \citet{delle2022common}.\footnote{Alternative estimation strategies include, among others, sequential least squares \citep{rodriguez2019multilevel} and sequential principal component approaches \citep{wang2010large}; see \citet{breitung2016analyzing} for a review of alternative estimation strategies.} The EM algorithm consists of two iterative steps that are repeated until convergence. First, in the Expectation step (E-step), the latent factors are estimated using the Kalman Smoother, based on an initial set of consistent parameter estimates. These estimated factors are then used in the Maximization step (M-step), where the model parameters are updated via linear projections, accounting for the uncertainty in the factor estimation \citep{doz2012quasi}. Convergence is assessed based on the likelihood derived from the Kalman filter in the E-step. Upon convergence, the algorithm provides quasi-maximum likelihood estimates for all model parameters. A final application of the Kalman Smoother yields consistent estimates of the latent factors; see \citet{barigozzi2022quasi} for a detailed discussion on the EM algorithm and its asymptotic properties. 

Initial estimates of all the parameters in the ML-DFM are obtained after estimating the loadings of all factors (pervasive and country-specific) via a top-down principal component (PC) procedure, which also provides initial estimates of the factors; see \citet{aastveit2016world}. Subsequently, these estimated factors are used to obtain initial estimates of the autoregressive parameters through separate least squares regressions for each factor; see Appendix \ref{app::modest} for a comprehensive explanation of the algorithm, its initialization, and comparisons with alternative estimation methods.

Finally, the Kalman Smoother also generates estimates of the Mean Squared Errors (MSEs) associated with the estimated factors. The MSEs of the factors delibered by the Kalman Smoother are obtained by assuming that the model specification is correct and that the estimated parameters coincide with the true model parameters. In order to incorporate parameter uncertainty into the MSE of the factors, we extend the subsampling procedure proposed by \cite{maldonado2021accurate} in the context of PC factor extraction, to the ML-DFM estimated using the EM algorithm; see Appendix \ref{app::modest}. We obtain the confidence regions for factors (and loadings) as proposed by Barigozzi and Luciani (2022) under the assumption of cross-sectionally uncorrelated idiosyncratic components; see Appendix \ref{app::modest} for a complete description of the procedure. Using these MSEs and assuming further normality, one can obtain the probabilistic contours of the factors needed to stress them. Note that the factors that enter into the FA-QR are both macroeconomic and financial factors. Even though these factors are extracted separately, they should be stressed based on their joint distribution. Consequently, when finding the joint contours of the factors, we assume that macroeconomic and financial factors are mutually independent; recall that their independence can be tested using the bootstrap procedure proposed by \cite{gonccalves2025bootstrap}.
 
\section{Data}
\label{sec::DA}

\begin{figure}[ht!]\caption{Quarterly GDP growth, and macroeconomic and financial series.}
\label{fig::dataC}
\centering \sc \footnotesize \smallskip
\setlength{\tabcolsep}{0.03\textwidth}
\vspace{-3pt}
\begin{tabular}{cc}
Germany: macro & Germany: financial  \\[3pt]
\includegraphics[width = 0.4\textwidth]{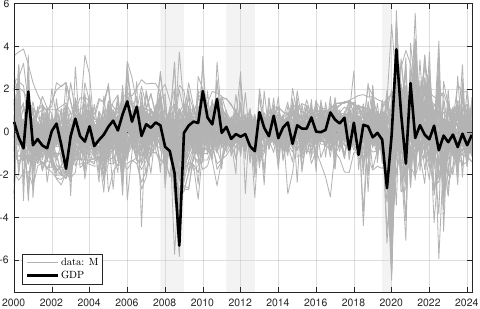} &
\includegraphics[width = 0.4\textwidth]{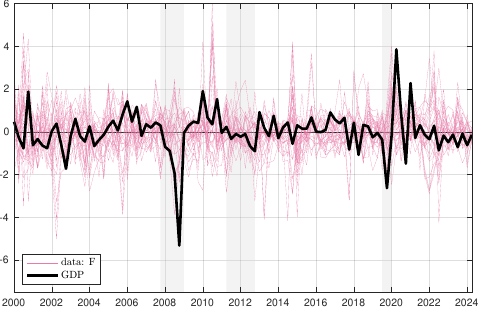} \\[3pt]
Spain: macro & Spain: financial  \\[3pt]
\includegraphics[width = 0.4\textwidth]{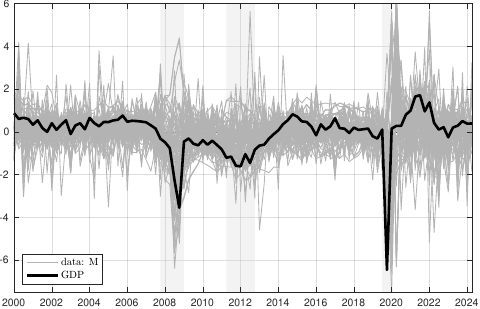} &
\includegraphics[width = 0.4\textwidth]{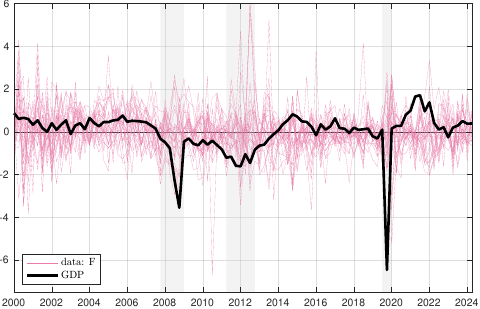} \\[3pt]
France: macro & France: financial  \\[3pt]
\includegraphics[width = 0.4\textwidth]{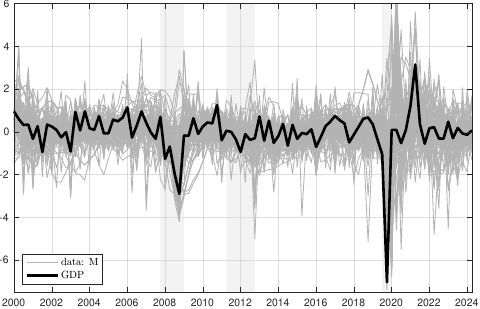} &
\includegraphics[width = 0.4\textwidth]{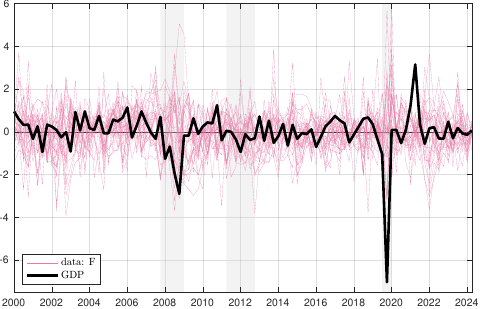} \\[3pt]
Italy: macro & Italy: financial  \\[3pt]
\includegraphics[width = 0.4\textwidth]{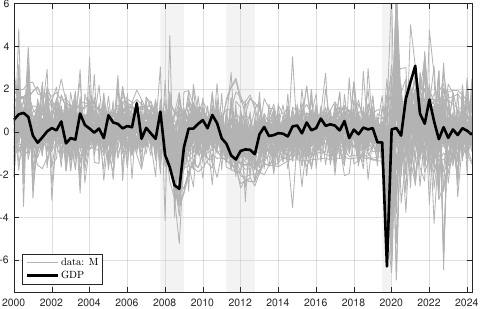} &
\includegraphics[width = 0.4\textwidth]{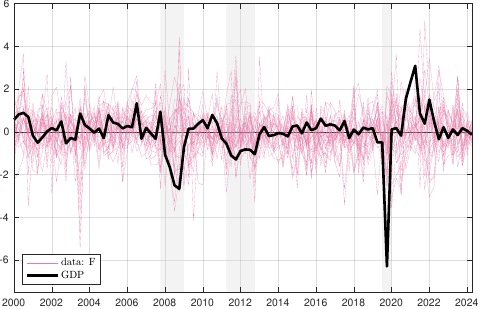} \\[3pt]
\end{tabular}
\begin{tabular}{p{0.9\textwidth}}\scriptsize \rm 
For each country, the figure plots GDP growth (black bold line) and macroeconomic (left panels) and financial (right panels) series observed quarterly from 2000Q2 to 2024Q3. Data are demeaned, standardized and cleaned for outliers, which explains the small drops in GDP growth during the Covid period.
\end{tabular}
\end{figure}

For each country, quarter-on-quarter GDP growth rates are observed quarterly from 2000Q2 to 2024Q3. The factors used to explain the one-step-ahead quantiles of the distribution of growth are extracted from the macroeconomic and financial variables contained in the EA-MD-QD database, as described by \citet{barigozzi2024large}.\footnote{Series are observed at the monthly level and aggregated at the quarterly level by simple quarterly averages.} Missing values due to ragged edges at the beginning/end of the sample are imputed using the EM algorithm procedure outlined by \citet{mccracken2016fred}. All variables are transformed so to achieve stationarity. In order to net out extreme observations, particularly due to the COVID-19 pandemic, outliers, which are defined as observations that deviate from the sample median by more than ten interquartile ranges, are replaced with a local median, computed using the nearest neighbourhood of 10 observations; see \citet{mccracken2020fred}. A complete description of the variables in the data base together with their corresponding transformations is reported in Appendix \ref{app::datades}. All series are finally demeaned and standardized. For each country, the macroeconomic and financial series used to extract the factors, alongside GDP growth, are plotted in Figure \ref{fig::dataC}.

\begin{figure}[ht!]\caption{Unconditional distribution of GDP growth in each country.}
\label{fig::uncD}
\centering \footnotesize \smallskip
\setlength{\tabcolsep}{0.03\textwidth}
\vspace{-3pt}
\begin{tabular}{c}
\includegraphics[width = 0.55\textwidth]{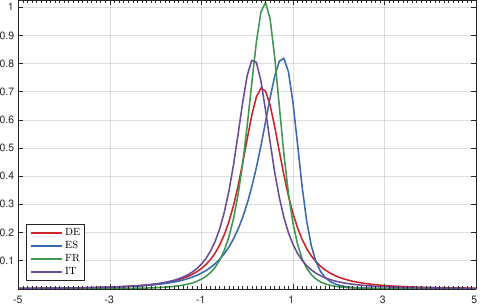}
\end{tabular}
\end{figure}

We estimate the unconditional distribution of GDP growth for each country by fitting a quantile regression with just a constant term, for $\tau =$ 0.25,0.50,0.75 and 0.95. Then, the estimated quantiles are used to fit the Skew-t distribution as described in Section \ref{sec::method}. Figure \ref{fig::uncD}, which plots the estimated unconditional density for each country, shows that the four countries considered differ not only in terms of their means, but also across the entire GDP growth distribution. We can observe that, although the average growth is larger in Spain, it also has the most skewed distribution to the left. The average growths are similar in Germany and France. Furthermore, both have approximately symmetric unconditional distributions. Finally, the average growth in Italy is smaller with an approximately symmetric distribution. The unconditional densities in all countries but France seem to have heavy tails; see also Table \ref{tab::uncD} that reports the parameters characterizing the estimated unconditional Skewed-t distributions. Specifically, Germany and Italy exhibit minimal skewness, as indicated by the parameter $\lambda$, with Italy showing the lowest mean GDP growth among the countries analysed. Despite this, both countries display the largest concentration of values in the tails, with their Skew-t distributions characterized by $\nu = 2$ degrees of freedom. In contrast, a different pattern emerges for France and Spain, particularly for the latter, which shows a notable degree of left skewness. Both countries also exhibit the highest average GDP growth, with France displaying the largest value of $\nu$, which suggests a relatively lower frequency of observations in the tails. Therefore, we align with the vast literature supporting the presence of heterogeneity between EA countries across different dimensions; see, for example, \citet{barigozzi2014euro}, \citet{gilchrist2018credit}, \citet{hoffmann2020financial}, and \citet{azqueta2023sources}.

\begin{table}[ht!]\caption{Unconditional Distribution of GDP growth: Estimated parameters of Skewed-t densities.}
\label{tab::uncD}
\centering \footnotesize \smallskip
\setlength{\tabcolsep}{0.03\textwidth}
\begin{tabular}[b]{c|cccc}
\hline
Country & $\mu$ & $\sigma^2$ & $\lambda$ & $\nu$ \\
\hline
DE & \hphantom{-}0.32 & \hphantom{-}0.50 & 0.02 & 2  \\
ES & \hphantom{-}1.05 & \hphantom{-}0.63 & -1.88 & 3  \\
FR & \hphantom{-}0.51 & \hphantom{-}0.39 & -0.55 & 4 \\
IT & \hphantom{-}0.14 & \hphantom{-}0.43 & -0.01 & 2 \\[3pt]
\end{tabular}
\begin{tabular}{p{.6\textwidth}}\scriptsize \rm
The table reports, for each country, the four parameters characterizing the Skew-t distributions: location ($\mu$), scale ($\sigma^2$), skewness ($\lambda$) and shape ($\nu$). 
\end{tabular}
\end{table}

\section{Estimating growth vulnerability in EA countries}
\label{sec::res}

In this section, after extracting the factors from the ML-DFM, we fit FA-QRs to estimate the quantiles of economic growth in each country. Finally, different scenarios of the corresponding conditional densities are obtained after stressing the factors.

\subsection{Extracting underlying factors}
\label{sbsec::resDFM}

The ML-DFM in \eqref{eq::mlDFM} is fitted separately to the macroeconomic and financial sets of variables to extract EA-wide and country-specific macroeconomic and financial factors, respectively. This modelling choice is backed by the bootstrap test of \citet{gonccalves2025bootstrap}, which indicates the absence of a common factor across the two datasets.\footnote{We run the test with $B=10000$ bootstrap replications, with a resulting $p$-value of $2 \times 10^{-4}$.} 
 After a preliminary analysis of the scree plots and the properties of the covariance matrices of the idiosyncratic noises, we determine for each data set one global factor ($r^{(\text{F})}=r^{(\text{M})}=1$). The number of country-specific macroeconomic factors is two per country($r^{(c,\text{M})} = 2$ $\forall c\in \mathcal{C}$), while the number of country-specific financial factors is one per country ($r^{(c,\text{F})} = 1$ $\forall c\in \mathcal{C}$). Consequently, the total number of factors is $r = 14$ with 9 macroeconomic factors and 5 financial factors; see Appendix \ref{app::modest} for a complete discussion of the model specification.

\begin{figure}[ht!]\caption{Global and country-specific factors.}
\label{fig::facts}
\centering \footnotesize \sc \smallskip
\setlength{\tabcolsep}{0.03\textwidth}
\vspace{-3pt}
\begin{tabular}{cc}
\multicolumn{2}{c}{Global factors} \\[3pt]
\scriptsize Macroeconomic & \scriptsize Financial  \\[3pt]
\includegraphics[width = 0.25\textwidth]{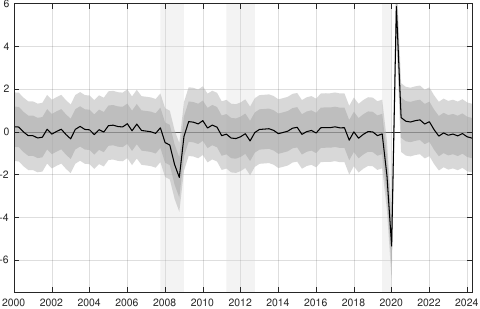} &
\includegraphics[width = 0.25\textwidth]{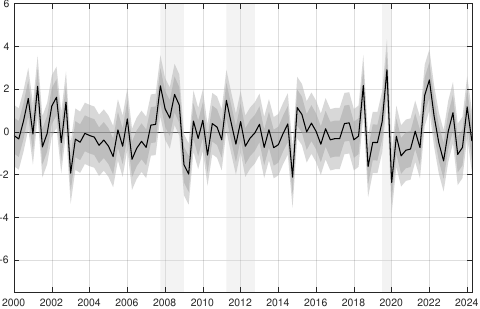} \\[6pt]
\end{tabular}
\begin{tabular}{ccc}
\multicolumn{3}{c}{Country-specific factors: Germany} \\[3pt]
\scriptsize Macroeconomic (1) & \scriptsize Macroeconomic (2) & \scriptsize Financial  \\[3pt]
\includegraphics[width = 0.25\textwidth]{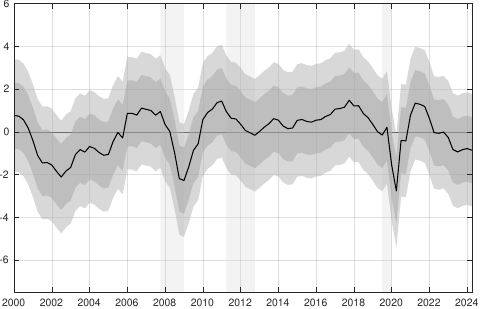} &
\includegraphics[width = 0.25\textwidth]{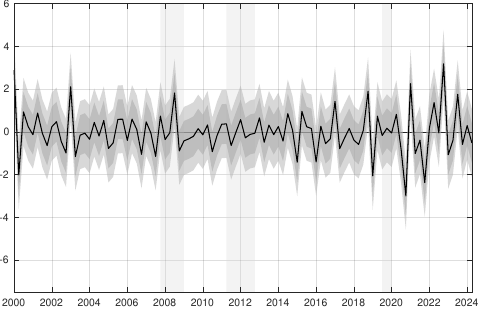} & 
\includegraphics[width = 0.25\textwidth]{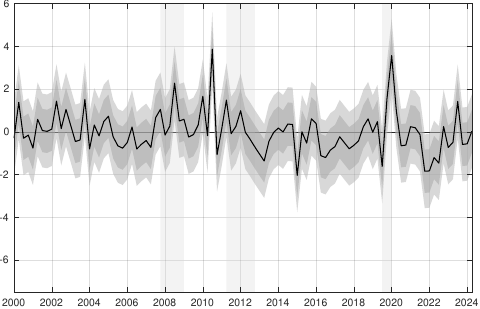} \\[6pt]
\multicolumn{3}{c}{Country-specific factors: Spain} \\[3pt]
\scriptsize Macroeconomic (1) & \scriptsize Macroeconomic (2) & \scriptsize Financial  \\[3pt]
\includegraphics[width = 0.25\textwidth]{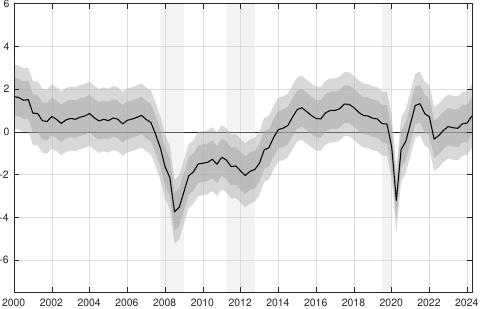} &
\includegraphics[width = 0.25\textwidth]{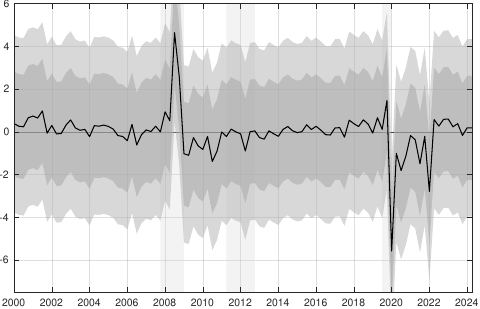} & 
\includegraphics[width = 0.25\textwidth]{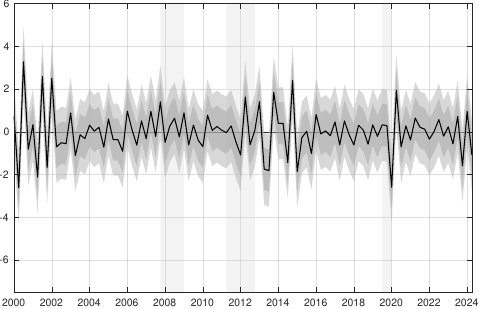} \\[6pt]
\multicolumn{3}{c}{Country-specific factors: France} \\[3pt]
\scriptsize Macroeconomic (1) & \scriptsize Macroeconomic (2) & \scriptsize Financial  \\[3pt]
\includegraphics[width = 0.25\textwidth]{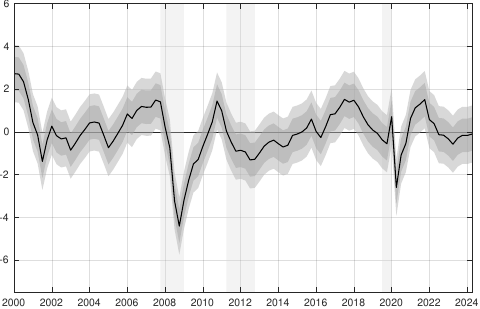} &
\includegraphics[width = 0.25\textwidth]{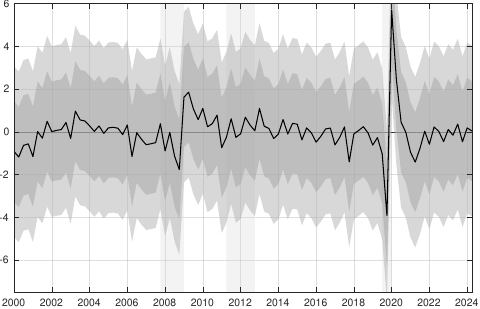} & 
\includegraphics[width = 0.25\textwidth]{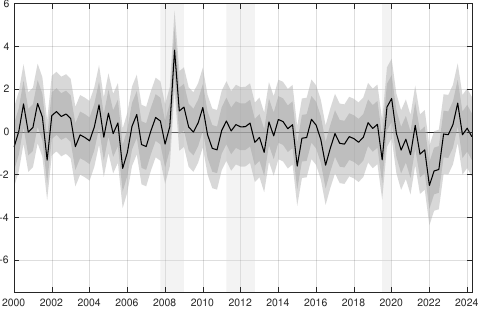} \\[6pt]
\multicolumn{3}{c}{Country-specific factors: Italy} \\[3pt]
\scriptsize Macroeconomic (1) & \scriptsize Macroeconomic (2) & \scriptsize Financial  \\[3pt]
\includegraphics[width = 0.25\textwidth]{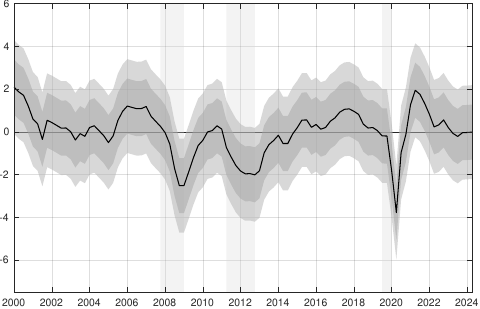} &
\includegraphics[width = 0.25\textwidth]{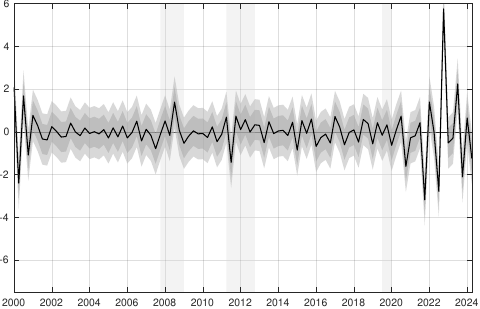} & 
\includegraphics[width = 0.25\textwidth]{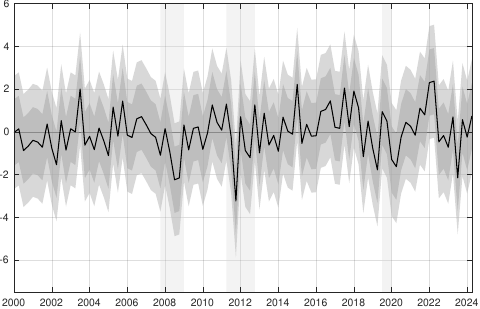} \\[6pt]
\end{tabular}
\begin{tabular}{p{.95\textwidth}}\scriptsize \rm
The figure plots the global and country-specific factors for each country (black solid lines), along with their 75\% and 95\% confidence bands (gray shaded areas). 
\end{tabular}
\end{figure}

The factors are extracted using the EM algorithm as outlined in Section \ref{sbsec::MLVDFM}. Figure \ref{fig::facts} plots the estimated global and country-specific macroeconomic and financial factors together with their corresponding 75\% and 90\% confidence bounds obtained from the Kalman Smoother. The first two panels plot the global factors, which are crucial for understanding the interdependence among countries. We can observe that the global macroeconomic factor evolves smoothly with big impacts of the global financial crisis and the COVID-19 pandemic, while the dynamic evolution of the global financial factor is more erratic. The remaining panels plot the country-specific macroeconomic and financial factors. These factors allow for the assessment of the relative position of each country in comparison to the aggregate EA-wide stance. The first country-specific macroeconomic factor is rather strong and shows the different patterns of the global crisis in each country. The second macroeconomic factor is rather erratic in Germany and Italy, while it jumps during the global financial crisis and the COVID19 pandemic in France and Spain. The magnitude of these factors is comparable to that of the global macroeconomic factor. It is remarkable the large uncertainty associated to the second macroeconomic factor, related to prices, in France and Spain. Finally, the local financial factors, which are comparable in magnitude to the global financial factor, are also rather erratic without showing any remarkable pattern.

\begin{figure}[ht!]\caption{Global and country-specific loadings.}
\label{fig::loads}
\centering \footnotesize \sc \smallskip
\setlength{\tabcolsep}{0.03\textwidth}
\vspace{-3pt}
\begin{tabular}{c}
\includegraphics[scale=1.5]{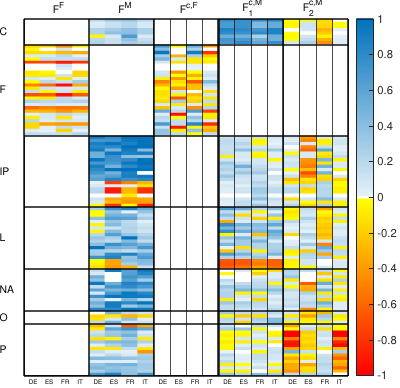}
\end{tabular}
\begin{tabular}{p{.95\textwidth}}\scriptsize \rm
The heatmap shows the magnitude of the estimated loadings of the global and country-specific factors for each country for financial (F) and macroeconomic variables, the latter grouped as: Confidence indices (C), Industrial Production (IP), Labour Market (L), National Accounts (NA), Others (O) and Prices (P). Bold vertical lines divide factors while thin lines divide countries within country-specific factors. By construction, the global financial and macroeconomic factors load on the sectoral variables across all countries, hence the absence of thin vertical separators.
\end{tabular}
\end{figure}

The loadings corresponding to the estimated factors are plotted in Figure \ref{fig::loads}. Consider first the results corresponding to the loadings of the macroeconomic factors. Note that the global factors loads on all variables across all countries but the variables corresponding to Confidence Indexes (C), some labour variables (Real Labour Productivity, Wages and Salaries, and Employer's Social Contribution, denote these variables as L$_1$), and all price variables (P). Remarkably, the first country specific macroeconomic factor loads in the variables in C and L$_1$. Furthermore, the second country-specific factor loads in all P variables. Therefore, while the factor underlying Industrial Production (IP), National Accounts (NA), Others (O) and the labour variables that are not in L$_1$ (denote these variables as L$_2$) is truly common across countries, the two factors underlying prices and confidence indexes, respectively, are country-specific. Furthermore, we can observe that the loadings of the global factor are positive for all countries and variables in IP, NA, O, and L$_2$ variables but for some IP variables. This last group of IP variables have negative weights in all countries but in Germany, in which they have zero weights. These results reinforce the presence of strong comovements across countries within the macroeconomic sector, aligning with the literature, which highlights the existence of international business cycles \citep{kose2003international,runstler2018business,oman2019synchronization}. Looking at the loadings of the first country-specific macroeconomic factor, linked to confidence indexes and productivity, we can observe that they are positive for the first group of variables while they are negative for the second. This second factor highlights the heterogeneity among countries linked to confidence and productivity, highlighting the impact of country-specific economic conditions \citep{runstler2018business,barigozzi2024large}. Finally, the second country-specific macroeconomic factor loads negatively in prices in Germany and Italy. However, the loadings of the price factor are close to zero in France and very mild in Spain. Once more, we observe an strong heterogeneity between countries with respect to the price factor.

Moving to the loadings of the financial factors, we can observe that the global factor loads uniformly in all countries with negative weights which are stronger on the stock market, while positive ones being mostly concentrated on long-term interest rates and exchange rates. However, the country-specific financial factors load on different variables in different countries.

Together, the common and country-specific factors offer a parsimonious yet comprehensive representation of the underlying macroeconomic and financial conditions in each country.

\subsection{GiS and GaR across Euro Area countries}
\label{sbsec::gargis}

In this section, we assess the power of global EA-wide and country-specific factors to predict the quantiles of economic growth in each country, and, consequently, its economic vulnerability. We also investigate whether their impact varies across EA countries, and estimate growth vulnerability not only in \enquote{normal} times but also when the factors are representative of a \enquote{stressed} economy.

Heterogeneity of the conditional growth distribution in each economy can arise either from the different impact of global EA-wide shocks affecting all countries or from the impact of country-specific macroeconomic and financial factors, which account for differences in sectoral behaviour across countries. The interplay between global and country-specific factors ultimately determines their overall impact on the distribution of economic growth and, consequently, on its vulnerability, measured by a extreme left quantile of this distribution. 

\begin{figure}[ht!]\caption{Estimated parameters of factor-augmented quantile regressions.}
\label{fig::coeffs}
\centering  \scriptsize \smallskip
\setlength{\tabcolsep}{0.005\textwidth}
\vspace{-3pt}
\begin{tabular}{ccc}
$\mu$ & $\phi$ & $\beta_1$: Global financial (F) factor\\[3pt]
\includegraphics[width = 0.33\textwidth]{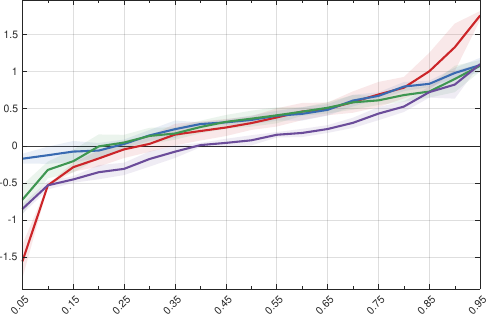} &
\includegraphics[width = 0.33\textwidth]{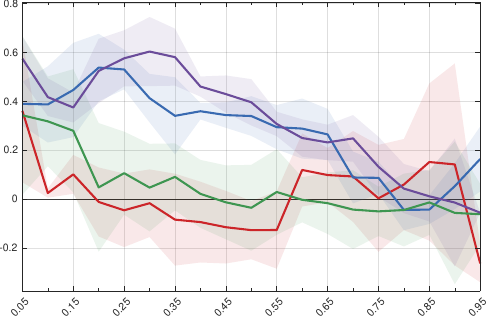} &
\includegraphics[width = 0.33\textwidth]{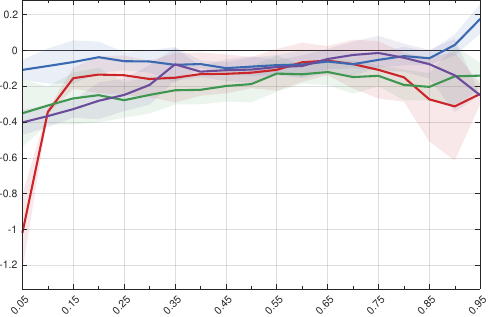} \\[3pt]
 $\beta_2$: Global macroeconomic (M) factor & $\beta_3$: Country-specific F factor & $\beta_4$: country-specific M factor (1)  \\[3pt]
\includegraphics[width = 0.33\textwidth]{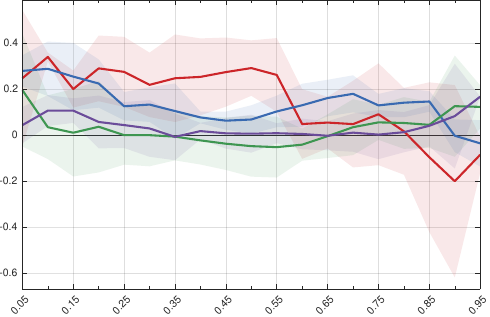} &
\includegraphics[width = 0.33\textwidth]{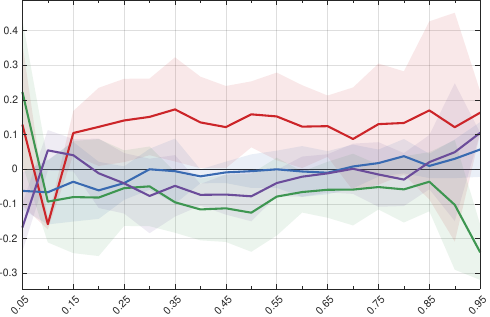} &
\includegraphics[width = 0.33\textwidth]{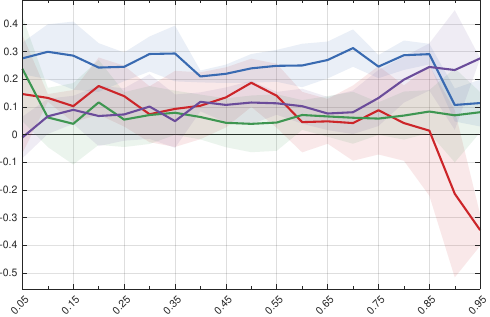} \\[3pt]
\end{tabular}
\begin{tabular}{ccc}
& $\beta_5$: Country-specific M factor (2) & \\[3pt]
& \includegraphics[width = 0.33\textwidth]{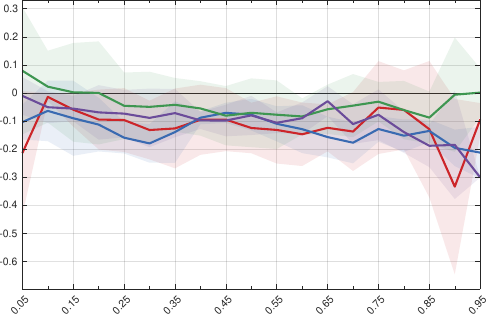} &\\[3pt]
\end{tabular}
\begin{tabular}{p{.95\textwidth}}\scriptsize \rm
The figure plots the estimated coefficients of the FA-QRs at each quantile $\tau$, from $\tau_{} = 0.05$ to $\tau = 0.95$, with steps of $0.05$, for each country: Germany (red solid lines), France (green solid lines), Spain (blue solid lines) and Italy (purple solid lines). Shades areas represent 95\% confidence bounds with the same colours for each country. 
\end{tabular}
\end{figure}

\noindent For each country, Figure \ref{fig::coeffs} plots the estimated parameters of the FA-QR models in (\ref{eq::Qreg}) together with their 95\% confidence bounds as a function of the quantile of growth for $\tau=0.05$ to 0.95 with steps of 0.05. The first conclusion from Figure \ref{fig::coeffs} is that, regardless of the particular country considered, the estimated factors play a significant role in shaping the distribution of GDP growth. This effect is stronger in the tails, particularly at the left tail. Second, it is remarkable the large and evident heterogeneity of conditional growth in the four economies analysed. Look first at the constant of the FA-QRs, $\mu$. They are all similar at the median, but for Italy, which has a clearly smaller parameter $\mu$. It is also remarkable that $\mu$ is clearly larger for Germany when looking at the right quantiles of the distribution of growth, implying a larger growth in expansions than that observed in any other economy. Also, there is a large heterogeneity in the level of growth in the left 0.05 quantile. The heterogeneity between the growth distribution in the four economies considered is even larger when looking at the dependence of this distribution with respect to past growth, measured by the parameter $\phi$. This parameter is hardly significant for the two largest economies, Germany and France, in which we only observe certain significant dependence of the extreme left quantiles. However, the conditional distribution of growth in Italy and Spain is clearly positively influenced by past growth, with this dependence decreasing as the quantile increases. For these two latter economies, past growth is more relevant in recessions than in expansions.

Finally, even more interesting is the heterogeneity observed in the dependence of growth with respect to the underlying macroeconomic and financial factors. Regardless of the economy, the global financial factor, $f_t^{(F)}$, is significant and negative. However, its influence is very strong in Germany and weak in Spain. The effect of the global macroeconomic factor is smaller in Italy and France and stronger in Germany. With respect to the country-specific factors, which represent different underlying mechanisms across countries, 
we can observe a very strong first macroeconomic factor, linked to confidence, in Spain while the country-specific financial factor is strong in France. 

To have a closer look at the impact of the factors on growth, Table \ref{tab::tabcoeff} reports the estimated parameters of the FA-QRs for some selected quantiles, in particular, the 5\%, 50\% and 95\% quantiles in each country, together with the corresponding $p$-values and $R^1$ coefficients. Given that one of the main interests of estimating the conditional density of growth is measuring vulnerability, we focus on the results for the 5\% quantile. 
 The financial factors are not relevant in Spain, with the country-specific factor not being even significant to explain the 5\% quantile of growth. However, the macroeconomic factors are very relevant in Spain; see that the $R^1$ coefficient is 0.40. However, in Germany France and Italy, the global macroeconomic factor is not significant while the global financial factor is very strong. These findings align with the literature, which highlights the interplay between local macro-financial conditions and EA-wide factors in shaping heterogeneous responses across countries; see \citet{cavallo2015common} and \citet{rathke2022similar}.
 
\begin{table}[ht!]
\caption{Estimated parameters of factor-augmented quantile regressions.}  
\label{tab::tabcoeff}
\vspace{-8pt}
\centering 
\small 
\vspace{5pt}

\resizebox{6.67in}{!}{ 
\begin{tabular}{c|ccc|ccc|ccc|ccc} 
\hline\hline 

& \multicolumn{ 3 }{ c | }{ \small Germany } 
& \multicolumn{ 3 }{ c | }{ \small Spain } 
& \multicolumn{ 3 }{ c | }{ \small France } 
& \multicolumn{ 3 }{ c  }{ \small Italy }  \\ \hline 
\footnotesize Coefficients 
& \footnotesize $\tau = .05$ &\footnotesize $\tau = .50$ &\footnotesize $\tau = .95$
& \footnotesize $\tau = .05$ &\footnotesize $\tau = .50$ &\footnotesize $\tau = .95$
& \footnotesize $\tau = .05$ &\footnotesize $\tau = .50$ &\footnotesize $\tau = .95$
& \footnotesize $\tau = .05$ &\footnotesize $\tau = .50$ &\footnotesize $\tau = .95$ \\ 
\hline
$\mu$
& -1.55 & \hphantom{-}0.30 & \hphantom{-}1.76 
& -0.17 & \hphantom{-}0.35 & \hphantom{-}1.09
& -0.72 & \hphantom{-}0.37 & \hphantom{-}1.08
& -0.85 & \hphantom{-}0.07 & \hphantom{-}1.10 \\[-3pt]
& \hphantom{-}\scriptsize (0.00) & \hphantom{-}\scriptsize (0.00) & \hphantom{-}\scriptsize (0.00)
& \hphantom{-}\scriptsize (0.00) & \hphantom{-}\scriptsize (0.00) & \hphantom{-}\scriptsize (0.00)
& \hphantom{-}\scriptsize (0.00) & \hphantom{-}\scriptsize (0.00) & \hphantom{-}\scriptsize (0.00)
& \hphantom{-}\scriptsize (0.00) & \hphantom{-}\scriptsize (0.00) & \hphantom{-}\scriptsize (0.00) \\
$\phi$
& \hphantom{-}0.36 & -0.12 & -0.26 
& \hphantom{-}0.39 & \hphantom{-}0.34 & \hphantom{-}0.16 
& \hphantom{-}0.34 & -0.04 & \hphantom{-}0.06
& \hphantom{-}0.58 & \hphantom{-}0.40 & -0.06 \\[-3pt]
& \hphantom{-}\scriptsize (0.01) & \hphantom{-}\scriptsize (0.03) & \hphantom{-}\scriptsize (0.04)
& \hphantom{-}\scriptsize (0.00) & \hphantom{-}\scriptsize (0.00) & \hphantom{-}\scriptsize (0.01)
& \hphantom{-}\scriptsize (0.01) & \hphantom{-}\scriptsize (0.44) & \hphantom{-}\scriptsize (0.35)
& \hphantom{-}\scriptsize (0.00) & \hphantom{-}\scriptsize (0.00) & \hphantom{-}\scriptsize (0.26) \\
$\beta^{(F)}$
& -1.01 & -0.13 & -0.24 
& -0.11 & -0.10 & \hphantom{-}0.18 
& -0.35 & -0.19 & -0.14 
& -0.40 & -0.11 & -0.25 \\[-3pt]
& \hphantom{-}\scriptsize (0.00) & \hphantom{-}\scriptsize (0.01) & \hphantom{-}\scriptsize (0.02)
& \hphantom{-}\scriptsize (0.01) & \hphantom{-}\scriptsize (0.02) & \hphantom{-}\scriptsize (0.00)
& \hphantom{-}\scriptsize (0.00) & \hphantom{-}\scriptsize (0.00) & \hphantom{-}\scriptsize (0.01)
& \hphantom{-}\scriptsize (0.00) & \hphantom{-}\scriptsize (0.01) & \hphantom{-}\scriptsize (0.00) \\
$\beta^{(M)}$
& \hphantom{-}0.25 & \hphantom{-}0.31 & -0.09 
& \hphantom{-}0.28 & \hphantom{-}0.07 & -0.04
& \hphantom{-}0.20 & -0.05 & \hphantom{-}0.12 
& \hphantom{-}0.04 & 0.06 & \hphantom{-}0.17 \\[-3pt]
& \hphantom{-}\scriptsize (0.09) & \hphantom{-}\scriptsize (0.00) & \hphantom{-}\scriptsize (0.22)
& \hphantom{-}\scriptsize (0.00) & \hphantom{-}\scriptsize (0.38) & \hphantom{-}\scriptsize (0.11)
& \hphantom{-}\scriptsize (0.10) & \hphantom{-}\scriptsize (0.24) & \hphantom{-}\scriptsize (0.08)
& \hphantom{-}\scriptsize (0.26) & \hphantom{-}\scriptsize (0.42) & \hphantom{-}\scriptsize (0.01) \\
$\beta^{(F,c)}$
& \hphantom{-}0.13 & \hphantom{-}0.15 & \hphantom{-}0.16 
& -0.06 & -0.01 & \hphantom{-}0.06
& \hphantom{-}0.22 & -0.13 & -0.24 
& -0.17 & -0.08 & \hphantom{-}0.11 \\[-3pt]
& \hphantom{-}\scriptsize (0.27) & \hphantom{-}\scriptsize (0.03) & \hphantom{-}\scriptsize (0.09)
& \hphantom{-}\scriptsize (0.07) & \hphantom{-}\scriptsize (0.00) & \hphantom{-}\scriptsize (0.10)
& \hphantom{-}\scriptsize (0.04) & \hphantom{-}\scriptsize (0.01) & \hphantom{-}\scriptsize (0.00)
& \hphantom{-}\scriptsize (0.01) & \hphantom{-}\scriptsize (0.05) & \hphantom{-}\scriptsize (0.02) \\
$\beta_1^{(M,c)}$
& \hphantom{-}0.15 & \hphantom{-}0.21 & -0.35 
& \hphantom{-}0.28 & \hphantom{-}0.24 & \hphantom{-}0.11 
& \hphantom{-}0.24 & \hphantom{-}0.04 & \hphantom{-}0.08 
& -0.01 & \hphantom{-}0.12 & \hphantom{-}0.28 \\[-3pt]
& \hphantom{-}\scriptsize (0.11) & \hphantom{-}\scriptsize (0.00) & \hphantom{-}\scriptsize (0.00)
& \hphantom{-}\scriptsize (0.00) & \hphantom{-}\scriptsize (0.00) & \hphantom{-}\scriptsize (0.00)
& \hphantom{-}\scriptsize (0.01) & \hphantom{-}\scriptsize (0.27) & \hphantom{-}\scriptsize (0.00)
& \hphantom{-}\scriptsize (0.40) & \hphantom{-}\scriptsize (0.01) & \hphantom{-}\scriptsize (0.07) \\
$\beta_2^{(M,c)}$
& -0.21 & -0.13 & -0.10 
& -0.10 & -0.07 & -0.21 
& \hphantom{-}0.08 & -0.07 & -0.01 
& -0.01 & -0.08 & -0.30 \\[-3pt]
& \hphantom{-}\scriptsize (0.03) & \hphantom{-}\scriptsize (0.01) & \hphantom{-}\scriptsize (0.20)
& \hphantom{-}\scriptsize (0.02) & \hphantom{-}\scriptsize (0.00) & \hphantom{-}\scriptsize (0.00)
& \hphantom{-}\scriptsize (0.26) & \hphantom{-}\scriptsize (0.11) & \hphantom{-}\scriptsize (0.41)
& \hphantom{-}\scriptsize (0.45) & \hphantom{-}\scriptsize (0.03) & \hphantom{-}\scriptsize (0.00) \\
\hline
$R^1$
& \hphantom{-}0.20 & \hphantom{-}0.06 & \hphantom{-}0.28 
& \hphantom{-}0.40 & \hphantom{-}0.35 & \hphantom{-}0.22
& \hphantom{-}0.24 & \hphantom{-}0.08 & \hphantom{-}0.17
& \hphantom{-}0.29 & \hphantom{-}0.14 & \hphantom{-}0.22\\

\hline\hline 
\end{tabular} 

}
\begin{tabular}{p{1\textwidth}}\scriptsize
The table reports the estimated parameters from the quantile regression together with their $p$-values in parenthesis, for the quantiles $\tau = .05$ and $\tau = .95$, along with the median ($\tau = .50$), for each country. For each quantile, the table also reports the $R^1$ of the regression.
\end{tabular}
\end{table}

 
All in all, the estimated parameters reported in Table \ref{tab::tabcoeff} show that the role of the common and country-specific macroeconomic and financial factors is rather different in the four largest economies of the EA considered. Overall, the global financial factor plays a pivotal role for all countries, while the global macroeconomic factors seems to be less relevant in light of the inclusion of local macroeconomic conditions. At the country-level, on the contrary, local macroeconomic conditions are important for all countries, while country-specific financial conditions prove to be less relevant.

\begin{figure}[ht!]
\caption{Smoothed skew-t densities.}
\label{fig::skewt}
\centering \sc \footnotesize \smallskip
\setlength{\tabcolsep}{0.03\textwidth}
\vspace{-3pt}
\begin{tabular}{cc}
Germany & Spain  \\[3pt]
\includegraphics[width = 0.415\textwidth]{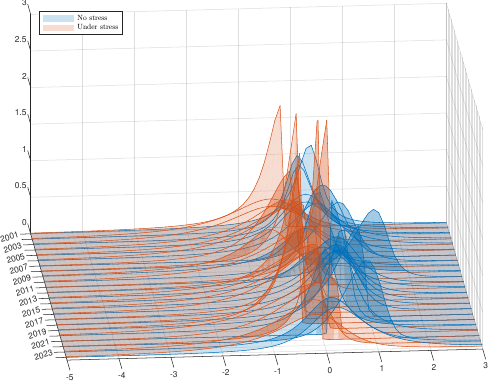} &
\includegraphics[width = 0.415\textwidth]{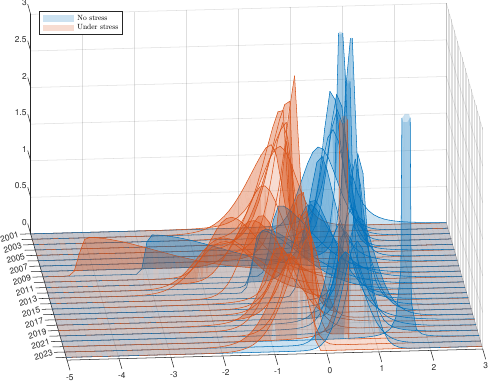} \\[3pt]
France & Italy  \\[3pt]
\includegraphics[width = 0.415\textwidth]{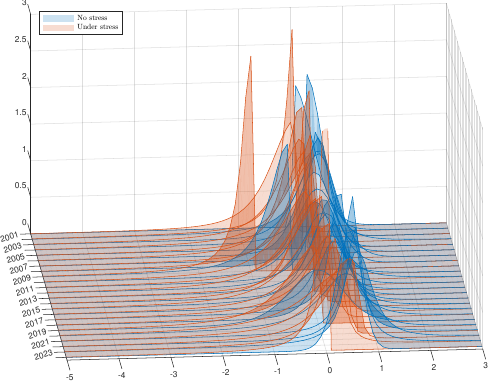} &
\includegraphics[width = 0.415\textwidth]{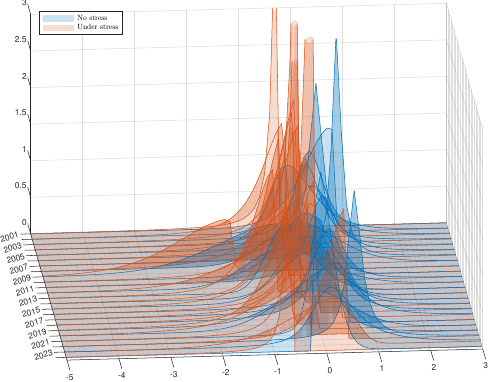} \\[3pt]
\end{tabular}
\begin{tabular}{p{.9\textwidth}}\scriptsize \rm
The figure plots, for each country, the smoothed Skew-t conditional densities, both without stress, i.e. when factors are fixed at their conditional mean (blue areas) and under stress, i.e. when the factors are jointly stressed at their 95\% level (red areas). Each area correspond to one time period (bottom-left axis), covering the last quarter of each year, starting from 2001Q1.
\end{tabular}
\end{figure}

Based on the estimated parameters of the FA-QR models reported in Table \ref{tab::tabcoeff}, we estimate the conditional growth density in each country under two different scenarios. The densities are first estimated as in \citet{adrian2019vulnerable}, when the factors are fixed at their conditional mean, i.e. under the "business as usual" scenario. Alternatively, the conditional growth densities are obtained as in \citet{gonzalez2024expecting}, when the factors are stressed at their 95th percentile level to minimize the 5\% quantile of growth. Figure \ref{fig::skewt} plots the conditional densities estimated under normal conditions (blue areas) and under stress (red areas). First, comparing the conditional densities obtained under the "business as usual" scenario with the unconditional densities plotted in Figure \ref{fig::uncD}, we can observe that, regardless of the particular country considered, the underlying factors explain movements not only of the mean and variance of the growth distribution but also its asymmetries, which are stronger in periods of economic crisis. Second, compared to normal times, the stressed densities exhibit two key characteristics: (i) they peak at lower levels of GDP growth and (ii) they show an increased skewness toward the right tail of the distribution. This pattern is consistent across all countries, although it appears to be more pronounced in Germany and Spain, particularly during periods of economic crisis, such as the Great Recession and the Covid-19 pandemic. It is also remarkable that, while the effect of stressing the underlying economic factors on the growth densities of France, Italy and Germany is rather marginal, the stressed growth densities corresponding to Spain are strongly different from those obtained under the "business as usual" scenario, showing the larger vulnerability of the Spanish economy.  

\begin{figure}[ht!]\caption{Estimated GaR and GiS}
\label{fig::gargis}
\centering \sc \footnotesize \smallskip
\setlength{\tabcolsep}{0.03\textwidth}
\vspace{-3pt}
\begin{tabular}{cc}
Germany & Spain  \\[3pt]
\includegraphics[width = 0.415\textwidth]{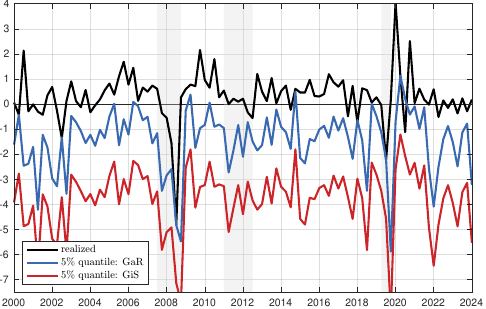} &
\includegraphics[width = 0.415\textwidth]{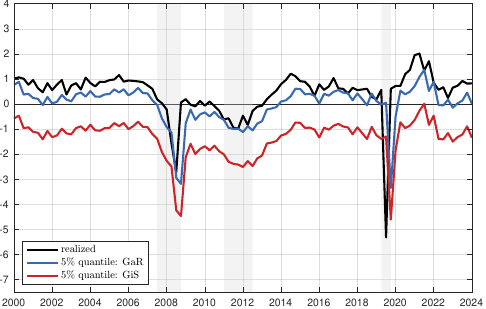} \\[3pt]
France & Italy  \\[3pt]
\includegraphics[width = 0.415\textwidth]{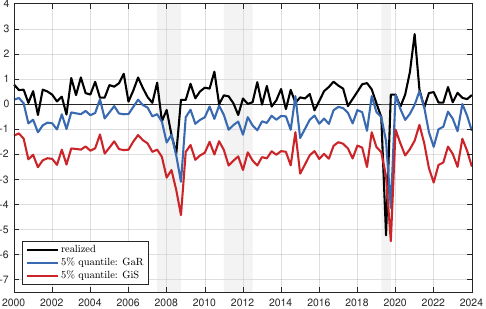} &
\includegraphics[width = 0.415\textwidth]{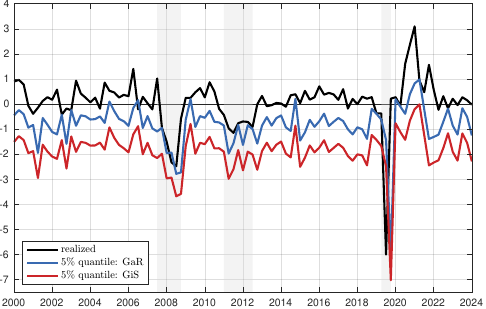} \\[3pt]
\end{tabular}
\begin{tabular}{p{.9\textwidth}}\scriptsize \rm
The figure plots observed GDP growth (black solid line), together with the estimated 5\% GaR (red solid line) and the 5\% GiS (blue solid line) when the underlying factors are jointly stressed at their 95\% probability level.
\end{tabular}
\end{figure}

Finally, we analyse the impact of the factors on both the GaR and the GiS of each country, defined as the 5th percentile of the corresponding GDP growth distribution obtanied under \enquote{business-as-usual} and stressed scenarios and computed as described in Section \ref{sbsec::gargis}. Figure \ref{fig::gargis} plots the (observed) GDP growth in each country, along with the corresponding GaR and GiS. Across all countries, stressed macroeconomic and financial conditions--both at the EA-wide and local levels--result in significantly higher growth vulnerability compared to the GaR measure. Specifically, under stressed conditions, the 5th percentile of GDP growth is, on average, between $\approx$3 percentage points lower for Germany and $1.1$ percentage points lower for Italy, relative to the corresponding percentile under the \enquote{business-as-usual} scenario.

The findings highlight the significance of macroeconomic and financial conditions, both at the EA-wide and country-specific levels, in shaping the distribution of GDP growth, particularly at the tails.\footnote{While in this paper we focus on vulnerable growth, the results from Table \ref{tab::tabcoeff} suggest that similar patterns emerge during periods of economic prosperity, when GDP growth is at the upper 95th percentile.} Moreover, the effects of common and country-specific economic conditions differ markedly across countries. For instance, some countries, such as Spain, are heavily influenced by local conditions, while others, particularly Italy, are predominantly impacted by EA-wide factors. These findings align with the literature, which highlights the interplay between local macro-financial conditions and EA-wide factors in shaping heterogeneous responses across countries; see \citet{cavallo2015common}  and \citet{rathke2022similar}.

Note that, despite the compelling evidence of cross-country heterogeneity in the impact of underlying economic factors on the distribution of growth, we are left with the question on the nature of the underlying sources of this heterogeneity. It is unclear how \enquote{stressed} economic conditions differ across countries and whether their effects can be driven by a small subset of factors being under stress, or whether stressing all factors is essential to the results observed thus far. Answering this important question will be the objective of the next subsection. 

\subsection{The role of macroeconomic and financial conditions in the Euro Area}
\label{sbsec::MF_EA}

In this subsection, we delve deeper into the nature of the shocks underlying the distribution of economic growth by exploring two key questions. First, we investigate whether stressed financial conditions, in addition to macroeconomic ones, are relevant for explaining economic growth vulnerability. Second, we examine whether the composition of the shocks driving the factors under stress has a significant impact.

First, to assess the role of financial shocks in the stressed growth distributions obtained above, we fix global and country-specific financial factors at their conditional mean and construct stressed scenarios for global and country-specific macroeconomic factors by stressing them at their 95\% probability level.
\footnote{It is important to note that this exercise is feasible as we are extracting factors separately from the set of financial and macroeconomic variables.
} We incorporate financial information into the model while isolating the effect of macroeconomic shocks. Table \ref{tab::tabdevF} reports the average through time of the absolute deviations of factors stressed at their 95\% level when minimizing the 5\% quantile of growth, with respect to their average in the "business-as-usual" scenario.\footnote{Since factors are identified only up to a sign, the sign of the factors alone is uninformative without a proper identification scheme, which allows for meaningful interpretation. Given that factors influence GDP growth quantiles through the quantile regression in \eqref{eq::Qreg}, the sign of the coefficients already indicates the role of each factor in shaping the conditional distribution of economic growth within each country.} The first column of Table \ref{tab::tabdevF} reports these deviations when all factors are jointly stressed while the second column reports the deviations when only the macroeconomic factors are stressed. The corresponding GiS are plotted in Figure \ref{fig::gwtshockM}. 
Once more, we can conclude that financial factors do not contribute much to stressing the distribution of growth in Spain. Figure \ref{fig::gwtshockM} shows that the GiS obtained when all factors are stressed and when only macroeconomic factors are stressed are nearly the same. Furthermore, the magnitude of the shocks to the macroeconomic factors is also very similar in both situations. However, in Germany, the shocks to the macroeconomic factors has to much larger when the financial factors are kept at their "business-as-usual" levels and, still the GiS is much smaller in absolute value in this last case. A similar situation can be observed in France and Italy with macroeconomic factors having much larger shocks and still the GiS being smaller in absoulte value when only the macroeconomic sector is under stress. This indicates that a higher degree of stress is necessary when focusing on a single sector alone. Stressing macroeconomic conditions alone, while leaving financial conditions unchanged, fails to produce the substantial levels of growth vulnerability observed under simultaneous stress of all factors. The case of Italy is remarkable with the 5\% quantile of growth obtained when only macroeconomic factors are stressed being not different from the 5\% quantile estimated in the "business-as-usual" scenario. This highlights the critical role of financial factors in amplifying adverse outcomes. Furthermore, the magnitude of this effect varies across countries, revealing different levels of exposure to financial conditions, both EA-wide and local. For instance, Germany and Spain appear more reliant on macroeconomic dynamics, whereas France and Italy--particularly the former--are significantly influenced by financial conditions. In these countries, neglecting financial shocks leads to a marked underestimation of downside risk to GDP, highlighting the importance of accounting for financial stress when assessing growth vulnerability.


\begin{table}[ht!]
\caption{Deviations of stressed factors from average values under alternative scenarios}  \label{tab::tabdevF}
\vspace{-8pt}
\centering 
\small 
\vspace{5pt}
\setlength{\tabcolsep}{0.01\textwidth}
\resizebox{6.6in}{!}{ 
\begin{tabular}{c c||c||c|c|c:c:c:c:c} 
\hline\hline
& & & 
\multicolumn{2}{c|}{\small \textbf{Joint Stress}} & 
\multicolumn{5}{c}{\small \textbf{Univariate Stress}} \\
\hline
& & 
\small \multirow{3}{*}{\textbf{Factors}} & 
\small \multirow{3}{*}{\begin{tabular}{c} Fin \\ \& \\ Macro \end{tabular}} & 
\small \multirow{3}{*}{\begin{tabular}{c} Only \\ Macro \end{tabular}} & 
\small \multirow{3}{*}{\begin{tabular}{c} Common \\ Fin \end{tabular}} & 
\small \multirow{3}{*}{\begin{tabular}{c} Common \\ Macro \end{tabular}} & 
\small \multirow{3}{*}{\begin{tabular}{c} Local \\ Fin \end{tabular}} & 
\small \multirow{3}{*}{\begin{tabular}{c} Local \\ Macro \\ 1 \end{tabular}} & 
\small \multirow{3}{*}{\begin{tabular}{c} Local \\ Macro \\ 2 \end{tabular}} \\
& & & & & & & & & \\
& & & & & & & & & \\
\hline
\multicolumn{1}{c|}{} \multirow{20}{*}{\STAB{\rotatebox[origin=c]{90}{\textbf{Country}}}} & 
\multirow{5}{*}{\STAB{\rotatebox[origin=c]{90}{\small DE}}} 
& $f^{\scaleto{\text{F}}{4pt}}$ & 1.36 & 0 & 2.48 & 0 & 0 & 0 & 0 \\
\multicolumn{1}{c|}{} 
& & $f^{\scaleto{\text{M}}{4pt}}$ & 1.41 & 1.03 & 0 & 2.57  & 0 & 0 & 0 \\
\multicolumn{1}{c|}{} 
& & $f^{\scaleto{(DE,\text{F})}{6pt}}$ & 0.59 & 0 & 0 & 0 & 2.70 & 0 & 0 \\
\multicolumn{1}{c|}{} 
& & $f_1^{\scaleto{(DE,\text{M})}{6pt}}$ & 2.58 & 2.87 & 0 & 0 & 0 & 4.77 & 0 \\ 
\multicolumn{1}{c|}{} 
& & $f_2^{\scaleto{(DE,\text{M})}{6pt}}$ & 0.83 & 1.36 & 0 & 0 & 0 & 0 & 2.77 \\
\cline{3-10}
\multicolumn{1}{c|}{} &
\multirow{5}{*}{\STAB{\rotatebox[origin=c]{90}{\small ES}}} 
& $f^{\scaleto{\text{F}}{4pt}}$ & 0.54 & 0 & 2.48 & 0 & 0 & 0 & 0 \\
\multicolumn{1}{c|}{} 
& & $f^{\scaleto{\text{M}}{4pt}}$ & 1.44 & 1.03 & 0 & 2.57 & 0 & 0 & 0 \\
\multicolumn{1}{c|}{} 
& & $f^{\scaleto{(ES,\text{F})}{6pt}}$ & 0.20 & 0 & 0 & 0 & 2.81 & 0 & 0 \\
\multicolumn{1}{c|}{} 
& & $f_1^{\scaleto{(ES,\text{M})}{6pt}}$ & 1.68 & 1.36 & 0 & 0 & 0 & 2.32 & 0 \\
\multicolumn{1}{c|}{} 
& & $f_2^{\scaleto{(ES,\text{M})}{6pt}}$ & 3.82 & 4.25 & 0 & 0 & 0 & 0 & 6.79 \\
\cline{3-10}
\multicolumn{1}{c|}{} & 
\multirow{5}{*}{\STAB{\rotatebox[origin=c]{90}{\small FR}}} 
& $f^{\scaleto{\text{F}}{4pt}}$ & 1.30 & 0 & 2.48 & 0 & 0 & 0 & 0 \\
\multicolumn{1}{c|}{} 
& & $f^{\scaleto{\text{M}}{4pt}}$ & 0.92 & 1.13 & 0 & 2.57 & 0 & 0 & 0 \\
\multicolumn{1}{c|}{} 
& & $f^{\scaleto{(FR,\text{F})}{6pt}}$ & 1.89 & 0 & 0 & 0 & 3.61 & 0 & 0 \\
\multicolumn{1}{c|}{} 
& & $f_1^{\scaleto{(FR,\text{M})}{6pt}}$ & 1.05 & 1.31  & 0 & 0 & 0 & 1.23 & 0 \\
\multicolumn{1}{c|}{} 
& & $f_2^{\scaleto{(FR,\text{M})}{6pt}}$ & 3.60 & 4.64 & 0 & 0 & 0 & 0 & 4.46 \\
\cline{3-10}
\multicolumn{1}{c|}{} & 
\multirow{5}{*}{\STAB{\rotatebox[origin=c]{90}{\small IT}}} 
& $f^{\scaleto{\text{F}}{4pt}}$ & 1.42 & 0 & 2.48 & 0 & 0 & & 0 \\
\multicolumn{1}{c|}{} 
& & $f^{\scaleto{\text{M}}{4pt}}$ & 1.47 & 2.02 & 0 & 2.57 & 0 & 0 & 0 \\
\multicolumn{1}{c|}{} 
& & $f^{\scaleto{(IT,\text{F})}{6pt}}$ & 2.45 & 0 & 0 & 0 & 4.26 & 0 & 0 \\
\multicolumn{1}{c|}{} 
& & $f_1^{\scaleto{(IT,\text{M})}{6pt}}$ & 0.29 & 1.10 & 0 & 0 & 0 & 3.87 & 0 \\
\multicolumn{1}{c|}{} 
& & $f_2^{\scaleto{(IT,\text{M})}{6pt}}$ & 0.14 & 0.14 & 0 & 0 & 0 & 0 & 2.03 \\
\hline\hline
\end{tabular} 

}
\begin{tabular}{p{0.98\textwidth}}\scriptsize

For each country, the table reports the average absolute deviation of the factors stressed at their $95\%$ level when minimizing the 5\% quantile of growth, with respect to their mean at the "business-as-usual" scenario. We consider two different scenarios: (i) all factors are under stress (first column); and (ii) only macroeconomic factors are stressed (second column). Univariate shocks to each factor obtained as the 95\% quantile of their corresponding marginal distributions are reported in the second panel of the table.
\end{tabular}
\end{table}

\begin{figure}[ht!]\caption{GiS under different stress scenarios.}
\label{fig::gwtshockM}
\centering \sc \footnotesize \smallskip
\setlength{\tabcolsep}{0.03\textwidth}
\vspace{-3pt}
\begin{tabular}{cc}
Germany & Spain  \\[3pt]
\includegraphics[width = 0.415\textwidth]{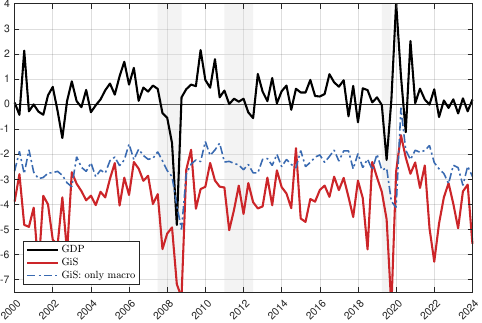} &
\includegraphics[width = 0.415\textwidth]{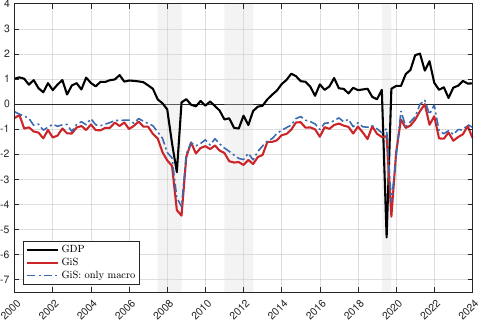} \\[3pt]
France & Italy  \\[3pt]
\includegraphics[width = 0.415\textwidth]{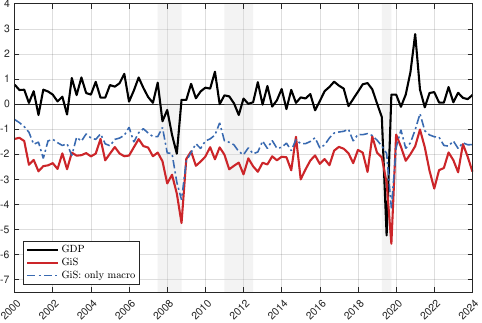} &
\includegraphics[width = 0.415\textwidth]{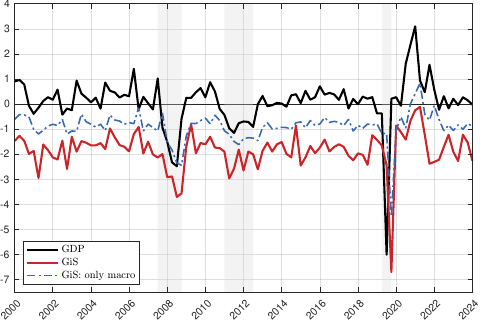} \\[3pt]
\end{tabular}
\begin{tabular}{p{0.9\textwidth}}\scriptsize \rm 
For each country, the figure plots observed GDP growth (black solid line), the estimated GiS when all factors are jointly stressed (red solid line) and when only macroeconomic factors are stressed (blue dashed line).
\end{tabular}
\end{figure}

\begin{figure}[ht!]\caption{Alternative stress scenarios: univariate sectoral shocks}
\label{fig::gwtshock}
\centering \sc \footnotesize \smallskip
\setlength{\tabcolsep}{0.03\textwidth}
\vspace{-3pt}
\begin{tabular}{cc}
Germany & Spain  \\[3pt]
\includegraphics[width = 0.415\textwidth]{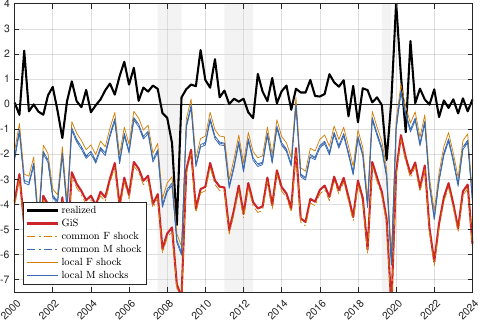} &
\includegraphics[width = 0.415\textwidth]{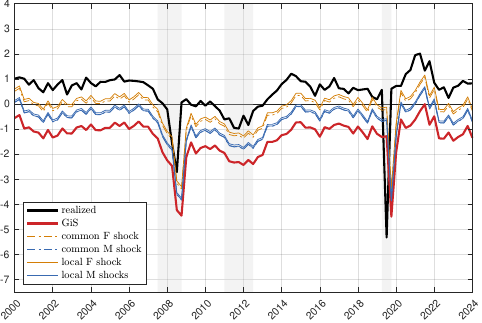} \\[3pt]
France & Italy  \\[3pt]
\includegraphics[width = 0.415\textwidth]{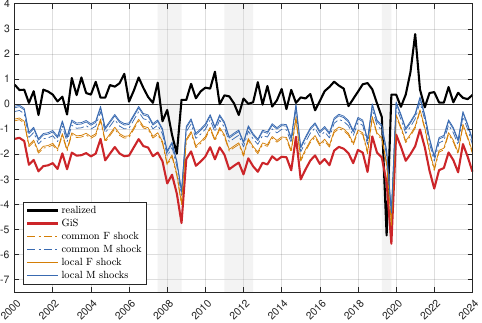} &
\includegraphics[width = 0.415\textwidth]{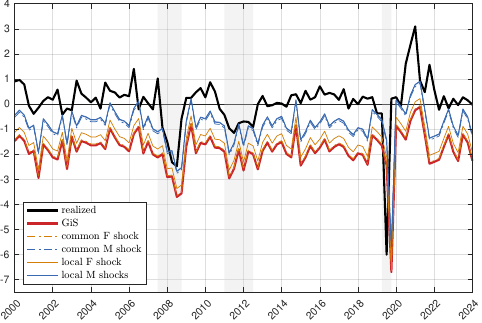} \\[3pt]
\end{tabular}
\begin{tabular}{p{0.9\textwidth}}\scriptsize \rm 
For each country, the figure plots observed GDP growth (black solid line), the estimated GiS when all factors are stressed (red solid line) and the GiS obtained in different scenarios; i) Only the global macroeconomic factor is stressed (dahsed blue line); ii) Only the global financial factor is stressed (dahsed orange line); iii) Only the country-specific macroeconomic factors are stressed (solid blue line); and iv) Only the country-specific financial factor is stressed (solid orange line).
\end{tabular}
\end{figure}

Second, we also assess the role of the composition of shocks driving the stressed factors by allowing for univariate shocks in the factors. We do so by considering the 95\% quantile of the corresponding marginal distributions of the common and country-specific factors; see Table \ref{tab::tabdevF} for the average magnitude of these univariate shocks. 
Through this exercise, we aim to determine whether the observed low economic growth under stressed conditions results from the joint depression of underlying macroeconomic and financial conditions, or whether very large shocks to individual sectors--either at the EA-wide level or at the country level--can explain the observed levels of vulnerability. Observe that the univariate shocks reported in Table \ref{tab::tabdevF} are much larger than those observed in their respective sectors when all factors are jointly stressed. 
 Figure \ref{fig::gwtshock} plots the estimated 5\% quantiles of growth when the factors are stressed one-by-one, together with the benchmark GiS, obtained when all factors are jointly stressed. We can observe that the high levels of economic growth vulnerability observed in Section \ref{sbsec::gargis} result from the overall economy--both in each economy and in the EA as a whole--being under pressure. Large idiosyncratic shocks to individual sectors, whether EA-wide or country-specific, are insufficient to fully account for the observed levels of economic growth vulnerability. An interesting exception is observed in the case of Italy, where large financial shocks, either global or specific, lead to a level of growth under stress comparable to that in the joint stress scenario. However, note that the implied financial shock is approximately 13 times larger than that observed when the economy is under general stress. This result could be explained by the high levels of debt observed in Italy, which make its financial conditions particularly sensitive to the EA-wide financial stance \citep[see, e.g., ][]{busetti2021time}. Furthermore, the effect of individual shocks--both common and country-specific--display significant heterogeneity across countries, reflecting the different impact of the macroeconomic and financial conditions on local markets.

This analysis yields two important insights. First, financial conditions under stress play a critical role in explaining GDP growth at its tails in all economies but Spain. Neglecting financial shocks leads to an underestimation of economic growth vulnerability, particularly during periods of stress when underlying economic conditions are strained.

Second, it is essential to account for both EA-wide and country-specific economic conditions as key drivers of the distribution of economic growth in each country. Large shocks to either of these factors have a sizeable impact on economic growth vulnerability although, alone, they are not able to fully explain the large negative values observed when the entire underlying economy is under stress. This distinction has important policy implications. Common shocks across EA countries often require coordinated responses, with macroprudential policies playing a key role in safeguarding financial stability and mitigating systemic risks. These policies can complement monetary policy actions, particularly in addressing vulnerabilities stemming from cross-border spillovers and shared financial exposures \citep{bussiere2021interaction}. Conversely, country-specific shocks are best managed through targeted macroprudential measures tailored to local conditions, among others \citep{claessens2015overview}. However, the interplay between these shocks--both across sectors and countries--can amplify systemic risks, underscoring the importance of an integrated policy approach. Coordinated macroprudential interventions, alongside monetary and fiscal tools, are essential to stabilize GDP growth and reduce the likelihood of severe negative economic outcomes when both EA-wide and country-specific conditions are under stress \citep{borio2014financial,cecchetti2016separation}.

\section{Conclusions}
\label{sec::conclude}

We study the heterogeneous dynamics of economic growth vulnerability across the four largest Euro Area economies under stressed macroeconomic and financial conditions. Using a multilevel dynamic factor model, we extract the pervasive EA-wide macroeconomic and financial  factors and semi-pervasive country-specific factors driving the distribution of GDP growth. This approach allows us to isolate the contributions of both EA-wide global and local country-specific macro-financial conditions, to understand their impact on growth vulnerabilities across countries. By focusing on the lower quantiles of the conditional distribution of GDP growth, we quantify vulnerability in scenarios where economic conditions deviate significantly from those observed during normal times, i.e. when the underlying factors are set at their average values.

The methodology adopted in this study offers a robust framework for analyzing growth vulnerabilities under stress. Factor-augmented quantile regression is a flexible yet parsimonious tool for capturing non-linear and asymmetric relationships between growth and economic conditions. This is complemented by the smooth estimation of conditional distributions using Skew-t densities to provide a detailed representation of downside risks to GDP growth. Finally, the integration of these elements with factor-based stress testing allows us to simulate adverse scenarios and assess growth vulnerability under systemic economic stress.

Our findings reveal distinct patterns of economic growth vulnerability, shaped by differences in exposure to EA-wide conditions and the strength of local dynamics. Specifically, countries heavily influenced by EA-wide conditions, such as Germany, and those driven by dominant country-specific dynamics, like Spain, display larger levels of growth vulnerability. The results of the construction of scenarios under stressed conditions reveal that, when only macroeconomic factors are stressed, while financial factors remain at their average levels, growth vulnerability is consistently smaller, underscoring the critical role of financial conditions in amplifying adverse economic outcomes. Furthermore, large sectoral shocks, whether global or country-specific, fail to fully explain the vulnerability observed under systemic stress, highlighting the interconnected and systemic nature of growth vulnerability across EA countries.

These results carry significant implications for policymaking. The evident heterogeneity in growth vulnerability underscores the necessity for a coordinated policy framework that integrates EA-wide macroprudential interventions with tailored country-specific measures. The interplay between common and local shocks emphasizes the importance of leveraging macroprudential tools alongside monetary and fiscal policies to address systemic risks and mitigate growth vulnerabilities effectively.

\newpage

\bibliographystyle{chicago}
\bibliography{refsEAG.bib}

\clearpage

\appendix

\gdef\thesection{\Alph{section}}
\gdef\thesubsection{\Alph{section}.\arabic{subsection}}
\gdef\thefigure{\Alph{section}\arabic{figure}}
\gdef\theequation{\Alph{section}\arabic{equation}}
\gdef\thetable{\Alph{section}\arabic{table}}

\setcounter{table}{0}
\setcounter{figure}{0}
\setcounter{equation}{0}
\setcounter{page}{1}

\gdef\thefootnote{(\roman{footnote})}

\section{Data Description}
\label{app::datades}

This appendix describes of the dataset, which is obtained from \cite{barigozzi2024large}. All series are retrieved from institutional sources. We indicate the source, the measurement unit and the seasonal adjustment, along with the transformation (if any) applied. Table \ref{gloss} presents a glossary to properly understand the data description presented in Tables \ref{tab:Germany} to \ref{tab:Spain}, which include a numeric indicator for each series (No), an alphabetical identifier (ID), a short description of the series (Series), the unit of measure (ID), the seasonal adjustment (SA), the frequency (F), the source (Source) and the transformation applied both in the light and in the heavy transformation settings (LT and HT, respectively). 

\begin{table}[H]
\caption{Glossary}
\label{gloss}
\resizebox{6.22in}{!}{%
\begin{tabular}{ l | l }
\hhline{==}
\textbf{Source} & \textbf{Unit} \\ \hhline{==}
EUR= Eurostat & CLV= Chain-linked volumes \\
OECD= Organization for Economic Co-operation and Development & $\text{CP}\_\text{MEUR}$ = Current Prices (Million\euro) \\
ECB = European Central Bank & 1000-ppl= Thousands of persons  \\
FRED = Federal Reserve Economic Data &  1000-U = Thousands of Units
\end{tabular}%
}
\end{table}
\begin{table}[H]
\resizebox{6.3in}{!}{%
\begin{tabular}{ l | l | l }
\hhline{===}
\textbf{Seasonal Adjustment} & \textbf{Frequency}  & \textbf{Transformation} \\ \hhline{===}
NSA = No Seasonal Adjustment & Q = Quarterly & 0 = No Transformation \\
SA = Seasonal Adjustment & M = Monthly & 1 = $100\times\log( x_t)$ \\
SCA = Seasonal and Calendar Adjustment & & 2 = $100\times \Delta\log(x_t)$  \\
MSA = Manual adjustment &  & 3 = $\Delta^2\log(x_t) $ \\
& & 4 = $\Delta x_t$ \\
& & 5 = $\Delta^2x_t$
\end{tabular}%
}
\end{table}
\newpage

\newgeometry{left=1.5cm,right=1.5cm,top=1cm,bottom=1.5cm}
\begin{table}[h!]
\caption{Data Description: Germany}
\label{tab:Germany}
\begin{footnotesize}
\resizebox{7.4in}{!}{%
\begin{tabular}{  c | c | l | c | c | c | c | c | c }
\hline
\hline
\textbf{No} & \textbf{ID} &	\textbf{Series} & \textbf{Unit} & \textbf{SA} & \textbf{F} & \textbf{Source} & \textbf{LT} & \textbf{HT}  \\ 
\hline
\hline
\multicolumn{9}{c}{\textbf{MACROECONOMIC INDICATORS}}\\
\hline
\hline
\multicolumn{9}{c}{(1) \textbf{National Accounts}}\\
\hline
1  & GDP    & Real Gross Domestic Product                                 			  & CLV(2015) & SCA & Q & EUR & 2 & 2 \\ 
2  & EXPGS  & Real Export Goods and services                              			  & CLV(2015) & SCA & Q & EUR & 2 & 2 \\ 
3  & IMPGS  & Real Import Goods and services                              			  & CLV(2015) & SCA & Q & EUR & 2 & 2 \\ 
4  & GFCE   & Real Government Final consumption expenditure               			  & CLV(2015) & SCA & Q & EUR & 2 & 2 \\ 
5  & HFCE   & Real Households consumption expenditure                     			  & CLV(2015) & SCA & Q & EUR & 2 & 2 \\
\hline
\hline
\end{tabular}%
}
\end{footnotesize}
\label{tab::dataDE}
\end{table}
\newpage
\newgeometry{left=1.5cm,right=1.5cm,top=1cm,bottom=1.5cm}
\begin{table}[h!]
\ContinuedFloat
\caption{Data Description: Germany}
\begin{footnotesize}
\resizebox{7.4in}{!}{%
\begin{tabular}{  c | c | l | c | c | c | c | c | c }
\hline
\hline
\textbf{No} & \textbf{ID} &	\textbf{Series} & \textbf{Unit} & \textbf{SA} & \textbf{F} & \textbf{Source} & \textbf{LT} & \textbf{HT}  \\ 
\hline
\hline
\multicolumn{9}{c}{(5) \textbf{Prices}}\\
\hline 
60 	& PPICAG   & Producer Price Index: Capital Goods 							   & 2015=100 & MSA & M & EUR  & 2 & 3 \\ 
61 	& PPICOG   & Producer Price Index: Consumer Goods 							   & 2015=100 & MSA & M & EUR  & 2 & 3 \\ 
62 	& PPIDCOG  & Producer Price Index: Durable Consumer Goods  					   & 2015=100 & MSA & M & EUR  & 2 & 3 \\ 
63 	& PPINDCOG & Producer Price Index: Non Durable Consumer Goods  				   & 2015=100 & MSA & M & EUR  & 2 & 3 \\ 
64 	& PPIING   & Producer Price Index: Intermediate Goods 						   & 2015=100 & MSA & M & EUR  & 2 & 3 \\ 
65 	& PPINRG   & Producer Price Index: Energy 									   & 2015=100 & MSA & M & EUR  & 2 & 3 \\
\hline
\hline
\end{tabular}%
}
\end{footnotesize}
\end{table}
\newpage
\begin{table}[h!]
\caption{Data Description: France}
\begin{footnotesize}
\resizebox{7.4in}{!}{%
\begin{tabular}{  c | c | l | c | c | c | c | c | c }
\hline
\hline
\textbf{No} & \textbf{ID} &	\textbf{Series} & \textbf{Unit} & \textbf{SA} & \textbf{F} & \textbf{Source} & \textbf{LT} & \textbf{HT}  \\ 
\hline
\hline
\multicolumn{9}{c}{\textbf{MACROECONOMIC INDICATORS}}\\
\hline
\hline
\multicolumn{9}{c}{(1) \textbf{National Accounts}}\\
\hline
1  & GDP    & Real Gross Domestic Product                                 			  & CLV(2015) & SCA & Q & EUR & 2 & 2 \\ 
2  & EXPGS  & Real Export Goods and services                              			  & CLV(2015) & SCA & Q & EUR & 2 & 2 \\ 
3  & IMPGS  & Real Import Goods and services                              			  & CLV(2015) & SCA & Q & EUR & 2 & 2 \\ 
4  & GFCE   & Real Government Final consumption expenditure               			  & CLV(2015) & SCA & Q & EUR & 2 & 2 \\ 
5  & HFCE   & Real Households consumption expenditure                     			  & CLV(2015) & SCA & Q & EUR & 2 & 2 \\
\hline
\hline
\end{tabular}%
}
\end{footnotesize}
\label{tab::dataFR}
\end{table}
\newpage
\newgeometry{left=1.5cm,right=1.5cm,top=1cm,bottom=1.5cm}
\begin{table}[h!]
\ContinuedFloat
\caption{Data Description: France}
\begin{footnotesize}
\resizebox{7.4in}{!}{%
\begin{tabular}{  c | c | l | c | c | c | c | c | c }
\hline
\hline
\textbf{No} & \textbf{ID} &	\textbf{Series} & \textbf{Unit} & \textbf{SA} & \textbf{F} & \textbf{Source} & \textbf{LT} & \textbf{HT}  \\ 
\hline
\hline
\multicolumn{9}{c}{(5) \textbf{Prices}}\\
\hline 
61 	& PPICAG   & Producer Price Index: Capital Goods 							   & 2015=100 & MSA & M & EUR  & 2 & 3 \\ 
62 	& PPICOG   & Producer Price Index: Consumer Goods 							   & 2015=100 & MSA & M & EUR  & 2 & 3 \\ 
63 	& PPIDCOG  & Producer Price Index: Durable Consumer Goods  					   & 2015=100 & MSA & M & EUR  & 2 & 3 \\ 
64 	& PPINDCOG & Producer Price Index: Non Durable Consumer Goods  				   & 2015=100 & MSA & M & EUR  & 2 & 3 \\ 
65 	& PPIING   & Producer Price Index: Intermediate Goods 						   & 2015=100 & MSA & M & EUR  & 2 & 3 \\ 
66 	& PPINRG   & Producer Price Index: Energy 									   & 2015=100 & MSA & M & EUR  & 2 & 3 \\
\hline
\hline
\end{tabular}%
}
\end{footnotesize}
\end{table}
\newpage
\begin{table}[h!]
\caption{Data Description: Italy}
\begin{footnotesize}
\resizebox{7.4in}{!}{%
\begin{tabular}{  c | c | l | c | c | c | c | c | c }
\hline
\hline
\textbf{N} & \textbf{ID} &	\textbf{Series} & \textbf{Unit} & \textbf{SA} & \textbf{F} & \textbf{Source} & \textbf{LT} & \textbf{HT}  \\ 
\hline
\hline
\multicolumn{9}{c}{\textbf{MACROECONOMIC INDICATORS}}\\
\hline
\hline
\multicolumn{9}{c}{(1) \textbf{National Accounts}}\\
\hline
1  & GDP    & Real Gross Domestic Product                                 			  & CLV(2015) & SCA & Q & EUR & 2 & 2 \\ 
2  & EXPGS  & Real Export Goods and services                              			  & CLV(2015) & SCA & Q & EUR & 2 & 2 \\ 
3  & IMPGS  & Real Import Goods and services                              			  & CLV(2015) & SCA & Q & EUR & 2 & 2 \\ 
4  & GFCE   & Real Government Final consumption expenditure               			  & CLV(2015) & SCA & Q & EUR & 2 & 2 \\ 
5  & HFCE   & Real Households consumption expenditure                     			  & CLV(2015) & SCA & Q & EUR & 2 & 2 \\
\hline
\hline
\end{tabular}%
}
\end{footnotesize}
\label{tab::dataIT}
\end{table}
\newpage
\newgeometry{left=1.5cm,right=1.5cm,top=1cm,bottom=1.5cm}
\begin{table}[h!]
\ContinuedFloat
\caption{Data Description: Italy}
\begin{footnotesize}
\resizebox{7.4in}{!}{%
\begin{tabular}{  c | c | l | c | c | c | c | c | c }
\hline
\hline
\textbf{N} & \textbf{ID} &	\textbf{Series} & \textbf{Unit} & \textbf{SA} & \textbf{F} & \textbf{Source} & \textbf{LT} & \textbf{HT}  \\ 
\hline
\hline
\multicolumn{9}{c}{(5) \textbf{Prices}}\\
\hline 
59 	& PPICAG   & Producer Price Index: Capital Goods 							   & 2015=100 & MSA & M & EUR  & 2 & 3 \\ 
60 	& PPICOG   & Producer Price Index: Consumer Goods 							   & 2015=100 & MSA & M & EUR  & 2 & 3 \\ 
61 	& PPIDCOG  & Producer Price Index: Durable Consumer Goods  					   & 2015=100 & MSA & M & EUR  & 2 & 3 \\ 
62 	& PPINDCOG & Producer Price Index: Non Durable Consumer Goods  				   & 2015=100 & MSA & M & EUR  & 2 & 3 \\ 
63 	& PPIING   & Producer Price Index: Intermediate Goods 						   & 2015=100 & MSA & M & EUR  & 2 & 3 \\ 
64 	& PPINRG   & Producer Price Index: Energy 									   & 2015=100 & MSA & M & EUR  & 2 & 3 \\
\hline
\hline
\end{tabular}%
}
\end{footnotesize}
\end{table}
\newpage
\begin{table}[h!]
\caption{Data Description: Spain}
\begin{footnotesize}
\resizebox{7.4in}{!}{%
\begin{tabular}{  c | c | l | c | c | c | c | c | c }
\hline
\hline
\textbf{N} & \textbf{ID} &	\textbf{Series} & \textbf{Unit} & \textbf{SA} & \textbf{F} & \textbf{Source} & \textbf{LT} & \textbf{HT}  \\ 
\hline
\hline
\multicolumn{9}{c}{\textbf{MACROECONOMIC INDICATORS}}\\
\hline
\hline
\multicolumn{9}{c}{(1) \textbf{National Accounts}}\\
\hline
1  & GDP    & Real Gross Domestic Product                                 			  & CLV(2015) & SCA & Q & EUR & 2 & 2 \\ 
2  & EXPGS  & Real Export Goods and services                              			  & CLV(2015) & SCA & Q & EUR & 2 & 2 \\ 
3  & IMPGS  & Real Import Goods and services                              			  & CLV(2015) & SCA & Q & EUR & 2 & 2 \\ 
4  & GFCE   & Real Government Final consumption expenditure               			  & CLV(2015) & SCA & Q & EUR & 2 & 2 \\ 
5  & HFCE   & Real Households consumption expenditure                     			  & CLV(2015) & SCA & Q & EUR & 2 & 2 \\
\hline
\hline
\end{tabular}%
}
\end{footnotesize}
\label{tab::dataES}
\end{table}

\newpage

\newgeometry{left=1.5cm,right=1.5cm,top=1cm,bottom=1.5cm}
\begin{table}[h!]
\ContinuedFloat
\caption{Data Description: Spain}
\label{tab:Spain}
\begin{footnotesize}
\resizebox{7.4in}{!}{%
\begin{tabular}{  c | c | l | c | c | c | c | c | c }
\hline
\hline
\textbf{N} & \textbf{ID} &	\textbf{Series} & \textbf{Unit} & \textbf{SA} & \textbf{F} & \textbf{Source} & \textbf{LT} & \textbf{HT}  \\ 
\hline
\hline
\multicolumn{9}{c}{(5) \textbf{Prices}}\\
\hline 
57 	& PPICAG   & Producer Price Index: Capital Goods 							   & 2015=100 & MSA & M & EUR  & 2 & 3 \\ 
58 	& PPICOG   & Producer Price Index: Consumer Goods 							   & 2015=100 & MSA & M & EUR  & 2 & 3 \\ 
59 	& PPIDCOG  & Producer Price Index: Durable Consumer Goods  					   & 2015=100 & MSA & M & EUR  & 2 & 3 \\ 
60 	& PPINDCOG & Producer Price Index: Non Durable Consumer Goods  				   & 2015=100 & MSA & M & EUR  & 2 & 3 \\ 
61 	& PPIING   & Producer Price Index: Intermediate Goods 						   & 2015=100 & MSA & M & EUR  & 2 & 3 \\ 
62 	& PPINRG   & Producer Price Index: Energy 									   & 2015=100 & MSA & M & EUR  & 2 & 3 \\
\hline
\hline
\end{tabular}%
}
\end{footnotesize}
\end{table}

\newpage

\newgeometry{left=2.5cm,right=2.5cm,top=2.5cm,bottom=2.5cm}
\section{Model and Estimation}
\label{app::modest}
\subsection{State-space representation}
\label{sbapp::ssR}
In this Section, we present the complete state-space form of the two-level dynamic factor model employed for estimation, obtained by stacking the data of each group in a single $n$-dimensional vector, where $N = \sum_{c\in \mathcal{C}}\sum_{s\in \mathcal{S}} N_{c,s}$.
Without loss of generality, let us order the groups by sector. Then, the model in compact form writes as:\\
\begin{subequations}
\begin{align}
\textbf{x}_t\ &=\ \boldsymbol{\Lambda} \mathbf{F}_t + \boldsymbol{\xi}_t; \quad \xi_{i,t}\sim \mathcal{N}(0,\sigma^2_{\xi_i})\label{sbeq::compobs}\\[5pt]
\textbf{F}_t\ &=\ \mathbf{A}\mathbf{F}_t + \mathbf{u}_t; \quad  \mathbf{u}_t \sim \mathcal{N}(\mathbf{0},\mathbf{Q})\label{sbeq::compst}
\end{align}
\end{subequations}
where:\\
\begingroup
\allowdisplaybreaks
\begin{subequations}
\begin{align}
\medmath{ \mathbf{x}_t}\ &\medmath{=\ 
\parT{  
\arraycolsep=2.5pt\def\arraystretch{0.95}
\begin{array}{cccccccc}
    \mathbf{x}_{t}^{\scaleto{(\text{DE},F)}{4pt}\prime} &
    \mathbf{x}_{t}^{\scaleto{(\text{ES},F)}{4pt}\prime} &
    \mathbf{x}_{t}^{\scaleto{(\text{FR},F)}{4pt}\prime} &
    \mathbf{x}_{t}^{\scaleto{(\text{IT},F)}{4pt}\prime} &
    \mathbf{x}_{t}^{\scaleto{(\text{DE},M)}{4pt}\prime} &
    \mathbf{x}_{t}^{\scaleto{(\text{ES},M)}{4pt}\prime} &
    \mathbf{x}_{t}^{\scaleto{(\text{FR},M)}{4pt}\prime} &
    \mathbf{x}_{t}^{\scaleto{(\text{IT},M)}{4pt}\prime} \\
    \end{array}
}'} \label{sbeq::compx}\\[15pt]   
\medmath{ \boldsymbol{\Lambda}}\ &\medmath{=\ 
\parT{
\arraycolsep=1.4pt\def\arraystretch{0.95}
\begin{array}{ll|ccccccccc}
     \bs{\Gamma}^{\scaleto{(\text{DE},F)}{4pt}} &
     \mathbf{0} & 
     \bs{\Phi}^{\scaleto{(\text{DE},F)}{4pt}} & 
     \mathbf{0} & 
     \mathbf{0} & 
     \mathbf{0} & 
     \mathbf{0} & 
     \mathbf{0} & 
     \mathbf{0} & 
     \mathbf{0} \\
     \bs{\Gamma}^{\scaleto{(\text{ES},F)}{4pt}} &
     \mathbf{0} & 
     \mathbf{0} & 
     \bs{\Phi}^{\scaleto{(\text{ES},F)}{4pt}} & 
     \mathbf{0} & 
     \mathbf{0} & 
     \mathbf{0} & 
     \mathbf{0} & 
     \mathbf{0} & 
     \mathbf{0}\\
     \bs{\Gamma}^{\scaleto{(\text{FR},F)}{4pt}} &
     \mathbf{0} &
     \mathbf{0} & 
     \mathbf{0} & 
     \bs{\Phi}^{\scaleto{(\text{FR},F)}{4pt}} & 
     \mathbf{0} & 
     \mathbf{0} & 
     \mathbf{0} & 
     \mathbf{0} & 
     \mathbf{0}\\
     \bs{\Gamma}^{\scaleto{(\text{IT},F)}{4pt}} &  
     \mathbf{0} &
     \mathbf{0} & 
     \mathbf{0} &
     \mathbf{0} & 
     \bs{\Phi}^{\scaleto{(\text{IT},F)}{4pt}} &
     \mathbf{0} & 
     \mathbf{0} & 
     \mathbf{0} & 
     \mathbf{0}\\
     \mathbf{0} &
     \bs{\Gamma}^{\scaleto{(\text{DE},M)}{4pt}} &         
     \mathbf{0} & 
     \mathbf{0} & 
     \mathbf{0} & 
     \mathbf{0} & 
     \bs{\Phi}^{\scaleto{(\text{DE},M)}{4pt}} & 
     \mathbf{0} &
     \mathbf{0} & 
     \mathbf{0}\\
     \mathbf{0} &
     \bs{\Gamma}^{\scaleto{(\text{ES},M)}{4pt}}  &        
     \mathbf{0} & 
     \mathbf{0} & 
     \mathbf{0} & 
     \mathbf{0} & 
     \mathbf{0} & 
     \bs{\Phi}^{\scaleto{(\text{ES},M)}{4pt}} &
     \mathbf{0} & 
     \mathbf{0} \\
     \mathbf{0} &
     \bs{\Gamma}^{\scaleto{(\text{FR},M)}{4pt}} &
     \mathbf{0} & 
     \mathbf{0} & 
     \mathbf{0} & 
     \mathbf{0} & 
     \mathbf{0} & 
     \mathbf{0} & 
     \bs{\Phi}^{\scaleto{(\text{FR},M)}{4pt}} &  
     \mathbf{0} \\
     \mathbf{0} & 
     \bs{\Gamma}^{\scaleto{(\text{IT},M)}{4pt}} &
     \mathbf{0} & 
     \mathbf{0} & 
     \mathbf{0} & 
     \mathbf{0} & 
     \mathbf{0} & 
     \mathbf{0} & 
     \mathbf{0} & 
     \bs{\Phi}^{\scaleto{(\text{IT},M)}{4pt}}\\
\end{array}}}\label{sbeq::compL}\\[15pt]   
\medmath{\mathbf{F}_t}\ &\medmath{=\
\parT{
\arraycolsep=2.5pt\def\arraystretch{0.95}
\begin{array}{cc|cccccccc}
\mathbf{f}_t^{\scaleto{(F)}{4pt}\prime} & \mathbf{f}_t^{\scaleto{(M)}{4pt}\prime} & 
\mathbf{f}_t^{\scaleto{(\text{DE},F)}{4pt}\prime} & 
\mathbf{f}_t^{\scaleto{(\text{ES},F)}{4pt}\prime} &
\mathbf{f}_t^{\scaleto{(\text{FR},F)}{4pt}\prime} & 
\mathbf{f}_t^{\scaleto{(\text{IT},F)}{4pt}\prime} &
\mathbf{f}_t^{\scaleto{(\text{DE},M)}{4pt}\prime} & 
\mathbf{f}_t^{\scaleto{(\text{ES},M)}{4pt}\prime} &
\mathbf{f}_t^{\scaleto{(\text{FR},M)}{4pt}\prime} & 
\mathbf{f}_t^{\scaleto{(\text{IT},M)}{4pt}\prime}
\end{array}}'}\label{sbeq::compF}\\[15pt]
\medmath{ \bs{\xi}_t}\ & \medmath{=\ 
\parT{  
\arraycolsep=2.5pt\def\arraystretch{0.95}
\begin{array}{cccccccc}
    \bs{\xi}_{t}^{\scaleto{(\text{DE},F)}{4pt}\prime} &
    \bs{\xi}_{t}^{\scaleto{(\text{ES},F)}{4pt}\prime} &
    \bs{\xi}_{t}^{\scaleto{(\text{FR},F)}{4pt}\prime} &
    \bs{\xi}_{t}^{\scaleto{(\text{IT},F)}{4pt}\prime} &
    \bs{\xi}_{t}^{\scaleto{(\text{DE},M)}{4pt}\prime} &
    \bs{\xi}_{t}^{\scaleto{(\text{ES},M)}{4pt}\prime} &
    \bs{\xi}_{t}^{\scaleto{(\text{FR},M)}{4pt}\prime} &
    \bs{\xi}_{t}^{\scaleto{(\text{IT},M)}{4pt}\prime} \\
\end{array}
}'} \label{sbeq::compE}\\[15pt]
\medmath{ \mathbf{A}}\ &\medmath{=\ 
\parT{  
\arraycolsep=2.5pt\def\arraystretch{0.95}
\begin{array}{cc|cccccccc}
\mathbf{A}^{\scaleto{(F)'}{4pt}} & \mathbf{0} & 
\mathbf{0} & \mathbf{0} & \mathbf{0} & 
\mathbf{0} & \mathbf{0} & \mathbf{0} & 
\mathbf{0} & \mathbf{0} \\
\mathbf{0} & \mathbf{A}^{\scaleto{(M)'}{4pt}} & 
\mathbf{0} & \mathbf{0} & \mathbf{0} & 
\mathbf{0} & \mathbf{0} & \mathbf{0} & 
\mathbf{0} & \mathbf{0}\\
\hline
\mathbf{0} & \mathbf{0} & 
\mathbf{A}^{\scaleto{(\text{DE},F)'}{4pt}} & \mathbf{0} & \mathbf{0} & 
\mathbf{0} & \mathbf{0} & \mathbf{0} &
\mathbf{0} & \mathbf{0}\\
\mathbf{0} & \mathbf{0} & 
\mathbf{0} & \mathbf{A}^{\scaleto{(\text{ES},F)'}{4pt}} & \mathbf{0} &
\mathbf{0} & \mathbf{0} & \mathbf{0} &
\mathbf{0} & \mathbf{0}\\
\mathbf{0} & \mathbf{0} & 
\mathbf{0} & \mathbf{0} & \mathbf{A}^{\scaleto{(\text{FR},F)'}{4pt}} & 
\mathbf{0} & \mathbf{0} & \mathbf{0} & 
\mathbf{0} & \mathbf{0}\\
\mathbf{0} & \mathbf{0} & 
\mathbf{0} & \mathbf{0} & \mathbf{0} & 
\mathbf{A}^{\scaleto{(\text{IT},F)'}{4pt}} & \mathbf{0} & \mathbf{0} &
\mathbf{0} & \mathbf{0}\\
\mathbf{0} & \mathbf{0} & 
\mathbf{0} & \mathbf{0} & \mathbf{0} & 
\mathbf{0} & \mathbf{A}^{\scaleto{(\text{DE},M)'}{4pt}} & \mathbf{0} &
\mathbf{0} & \mathbf{0}\\
\mathbf{0} & \mathbf{0} & 
\mathbf{0} & \mathbf{0} & \mathbf{0} & 
\mathbf{0} & \mathbf{0} & \mathbf{A}^{\scaleto{(\text{ES},M)'}{4pt}} &
\mathbf{0} & \mathbf{0}\\
\mathbf{0} & \mathbf{0} & 
\mathbf{0} & \mathbf{0} & \mathbf{0} & 
\mathbf{0} & \mathbf{0} & \mathbf{0} &
\mathbf{A}^{\scaleto{(\text{FR},M)'}{4pt}} & \mathbf{0}\\
\mathbf{0} & \mathbf{0} & 
\mathbf{0} & \mathbf{0} & \mathbf{0} & 
\mathbf{0} & \mathbf{0} & \mathbf{0} & 
\mathbf{0} & \mathbf{A}^{\scaleto{(\text{IT},M)'}{4pt}}\\
\end{array}
}} \label{sbeq::compA} \\[15pt]   
\medmath{\mathbf{u}_t}\ &\medmath{=\
\parT{
\arraycolsep=2.5pt\def\arraystretch{0.95}
\begin{array}{cc|cccccccc}
\mathbf{u}_t^{\scaleto{(F)}{4pt}\prime} & 
\mathbf{u}_t^{\scaleto{(M)}{4pt}\prime} & 
\mathbf{u}_t^{\scaleto{(\text{DE},F)}{4pt}\prime} & 
\mathbf{u}_t^{\scaleto{(\text{ES},F)}{4pt}\prime} &
\mathbf{u}_t^{\scaleto{(\text{FR},F)}{4pt}\prime} & 
\mathbf{u}_t^{\scaleto{(\text{IT},F)}{4pt}\prime} &
\mathbf{u}_t^{\scaleto{(\text{DE},M)}{4pt}\prime} & 
\mathbf{u}_t^{\scaleto{(\text{ES},M)}{4pt}\prime} &
\mathbf{u}_t^{\scaleto{(\text{FR},M)}{4pt}\prime} & 
\mathbf{u}_t^{\scaleto{(\text{IT},M)}{4pt}\prime}
\end{array}}'}\label{sbeq::compU}
\end{align}\\
and:
\begin{equation}
\mathbf{A}^{s}\ =\ 
\parT{
\begin{array}{ccccc}
\mathbf{A}_1^{(s)} & \mathbf{A}_2^{(s)} & \multicolumn{2}{c}{\ldots}  & \mathbf{A}_{p_s}^{(s)} \\
\mathbf{I} & \mathbf{0} & \multicolumn{2}{c}{\ldots}  &  \mathbf{0}\\
\mathbf{0} & \mathbf{I} & \multicolumn{2}{c}{\ldots}  & \mathbf{0}\\
\vdots & \vdots & \multicolumn{2}{c}{\ddots}  & \vdots \\
\mathbf{0} & \mathbf{0} & \ldots & \mathbf{I} & \mathbf{0}
\end{array}
} \hspace{20pt} \mathbf{A}^{(c,s)}\ =\ \parT{
\begin{array}{ccccc}
\mathbf{A}_1^{(c,s)} & \mathbf{A}_2^{(c,s)} & \multicolumn{2}{c}{\ldots}  & \mathbf{A}_{p_s}^{(c,s)} \\
\mathbf{I} & \mathbf{0} & \multicolumn{2}{c}{\ldots}  &  \mathbf{0}\\
\mathbf{0} & \mathbf{I} & \multicolumn{2}{c}{\ldots}  & \mathbf{0}\\
\vdots & \vdots & \multicolumn{2}{c}{\ddots}  & \vdots \\
\mathbf{0} & \mathbf{0} & \ldots & \mathbf{I} & \mathbf{0}
\end{array}
}
\label{sbeq::compA}
\end{equation}
\end{subequations}
\endgroup\\
are the companion form representation of the transition matrices for $s\in\mathcal{S}$ and $c\in\mathcal{C}$.

Finally, let $\bs{\Gamma}^{(s)} = (\bs{\Gamma}^{\scaleto{(\text{DE},s)}{6pt}\prime}, \bs{\Gamma}^{\scaleto{(\text{ES},s)}{6pt}\prime}, \bs{\Gamma}^{\scaleto{(\text{FR},s)}{6pt}\prime}, \bs{\Gamma}^{\scaleto{(\text{IT},s)}{6pt}\prime})' = (\bs{\gamma}_1^{(s)},\ldots, \bs{\gamma}_{n_s}^{(s)})'$ denote the loadings of the pervasive factors for sector $s\in \mathcal{S}$. 

\subsection{EM algorithm}
\label{sbapp:EM}

\subsubsection{Initialization}
\label{sbsbapp::initEM}

To estimate the model in Equations \eqref{sbeq::compobs}-\eqref{sbeq::compst}, we need to initialize all the parameters and states in the model. To do so, we employ the two step (top-down) PC approach \citep{aastveit2016world,breitung2016analyzing} to extract the estimated factors and the corresponding loadings, and we then use these estimated factors to obtain an estimate of the transition parameters in Equation \eqref{sbeq::compst}.

Specifically, let $\mathbf{X}^{(s)} = (\mathbf{x}_1^{(s)},\ldots,\mathbf{x}_T^{(s)})'$ be the $T\times N_s$ vector of data for sector $s\in\mathcal{S}$, pooling all countries. Without loss of generality, let $\mathbf{X}^{(s)}$ be standardized so to have mean zero and unit variance.
Given the data, we can estimate the global factor by means of principal components (PC) on the $N_s\times N_s$ sample variance-covariance matrix, say $\widehat{\boldsymbol{\Gamma}}^{X^s} = \frac{\mathbf{X}^{(s)\prime}\mathbf{X}^{(s)}}{T}$.

Following \citet{barigozzi2022estimation}, the least-squares estimation of the common factors within each sector, $\mathbf{f}^{(s)}$, and the corresponding loadings, $\bs{\Gamma}^{(s)}$, is given by:\\
\begin{equation}
\label{eq::LSsol}
\min_{\ubar{\mathbf{F}}^{(s)},\ubar{\bs{\Gamma}}^{(s)}}\ \frac{1}{N_s T}\sum_{t=1}^T \parT{\mathbf{X}^{(s)} - \hspace{-1.2pt}\ubar{\hspace{1.2pt}\mathbf{F}}^{(s)}\hspace{-1.2pt}\ubar{\hspace{1.2pt}\bs{\Gamma}}^{(s)\prime}}'\parT{\mathbf{X}^{(s)} - \hspace{-1.2pt}\ubar{\hspace{1.2pt}\mathbf{F}}^{(s)}\hspace{-1.2pt}\ubar{\hspace{1.2pt}\bs{\Gamma}}^{(s)\prime}}
\end{equation}\\
where $\hspace{-1.2pt}\ubar{\hspace{1.2pt}\mathbf{F}}^{(s)} = (\hspace{1pt}\ubar{\hspace{-1pt}\mathbf{f}}_1^{(s)},\ldots, \hspace{1pt}\ubar{\hspace{-1pt}\mathbf{f}}_T^{(s)})'$.
Under the identification conditions by \citet{forni2009opening}, i.e. $N_s^{-1}\hspace{-1.2pt}\ubar{\hspace{1.2pt}\bs{\Gamma}}^{(s)\prime}\hspace{-1.2pt}\ubar{\hspace{1.2pt}\bs{\Gamma}}^{(s)}$ diagonal, we solve Equation \eqref{eq::LSsol} by $\hspace{-1.2pt}\ubar{\hspace{1.2pt}\bs{\Gamma}}^{(s)}$, to obtain the PC estimator:\\
\begin{equation}
\label{eq::Ls}
\widehat{\bs{\Gamma}}^{(s),(0)}\ =\ \widehat{\mathbf{V}}^{\scaleto{\mathbf{X}^{(s)}}{5pt}}\parT{\widehat{\mathbf{M}}^{\scaleto{\mathbf{X}^{(s)}}{5pt}}}^{\frac{1}{2}}
\end{equation}\\
where $\widehat{\mathbf{V}}^{\scaleto{\mathbf{X}^{(s)}}{5pt}}$ is the matrix of eigenvectors corresponding to the $r_s-$ largest eigenvalues, $\widehat{\mathbf{M}}^{\scaleto{\mathbf{X}^{(s)}}{5pt}}$, of the sample variance-covariance matrix of the data. Given the estimated loadings, we obtain an estimate of the factor by projection onto matrix of data, that is:\\
\begin{equation}
\label{eq::Fs}
\widehat{\mathbf{F}}^{(s),(0)}\ =\ \mathbf{X}^{(s)}\widehat{\bs{\Gamma}}^{(s)}\parT{\widehat{\bs{\Gamma}}^{(s)\prime}\widehat{\bs{\Gamma}}^{(s)}}^{-1}\ =\ \mathbf{X}^{(s)}\widehat{\mathbf{V}}^{\scaleto{\mathbf{X}^{(s)}}{5pt}}\parT{\widehat{\mathbf{M}}^{\scaleto{\mathbf{X}^{(s)}}{5pt}}}^{-\frac{1}{2}} 
\end{equation}\\
Given the estimated factors in each sector, we still have to estimate the local factors within each country-sector block. Let $\overline{\mathbf{X}}^{(s)} = \mathbf{X}^{(s)} - \widehat{\mathbf{F}}^{(s),(0)}\widehat{\bs{\Gamma}}^{(s),(0)\prime}$ be the matrix of data for sector $s\in\mathcal{S}$ net of the comovements explained by the global factor (i.e., common to all countries) within the sector. We can repeat the procedure outlined above to obtain an estimate of the local factors within each country-sector subgroup.
For each country-sector pair, the estimated loadings are given by:\\
\begin{equation}
\widehat{\bs{\Phi}}^{(s),(0)}\ =\ \widehat{\mathbf{V}}^{\scaleto{\overline{\mathbf{X}}^{(s)}}{5pt}}\parT{\widehat{\mathbf{M}}^{\scaleto{\overline{\mathbf{X}}^{(s)}}{5pt}}}^{\frac{1}{2}}
\end{equation}
\subsubsection{E-step}
\label{sbsbapp:Estep}
Let $k = 1,\ldots, k_{\max}$ denote the indicator for the current iteration of the EM algorithm, and let $\widehat{\boldsymbol{\theta}}^{(k-1)}$ denote the vector of parameters estimated at iteration $k-1$, while $\mathbf{F}_{0\giv 0}^{(k-1)}$ and $\mathbf{P}_{0\giv 0}^{(k-1)}$ denote the corresponding initial estimate of factor and their conditional variance. Then, conditional on these quantities, in the E-step we run the Kalman Filter and Smoother to obtain updated estimates the factors and their variance, i.e. $\mathbf{F}_{t\giv T}^{(k)}$ and $\mathbf{P}_{t\giv T}^{(k)}$, along with the conditional covariance across states, denoted as $\mathbf{P}_{t,t-1\giv T}^{(k)}$.

\subsubsection{M-step}
\label{sbsbapp:Mstep}

In the M-step, we maximize the expected log-likelihood with respect to the parameters.\\

\begin{enumerate}
\item \textbf{Factor Loadings}:
\begin{align*}
\widehat{\bs{\gamma}}_i^{(s),(k+1)}\ &=\ \parT{\sum_{t=1}^T x_{i,t}^{(s)}\mathbf{f}_{t\giv T}^{(s),(k)}}\parT{\sum_{t=1}^T \mathbf{f}_{t\giv T}^{(s),(k)}\mathbf{f}_{t\giv T}^{(s),(k)\prime} + \mathbf{P}_{t\giv T}^{\mathbf{f}^{(s)},(k)}}^{-1} \\[5pt]
\widehat{\bs{\phi}}_i^{(c,s),(k)}\ &=\ \parT{\sum_{t=1}^T x_{i,t}^{(c,s)}\mathbf{f}_{t\giv T}^{(c,s),(k)}}\parT{\sum_{t=1}^T \mathbf{f}_{t\giv T}^{(c,s),(k)}\mathbf{f}_{t\giv T}^{(c,s),(k)\prime} + \mathbf{P}_{t\giv T}^{\mathbf{f}^{(c,s)},(k)}}^{-1} \\[5pt]
\end{align*}
stacking the loadings, we denote the estimated loadings $\widehat{\bs{\Lambda}}^{(k+1)}$.
\item \textbf{Idiosyncratic variances}:
\begin{align*}
\widehat{\sigma}^{2,(k+1)}_{\xi_i}\ =\ \frac{1}{T}\sum_{t=1}^T\parQ{\parT{x_{i,t} - \widehat{\bs{\lambda}}_i^{(k+1)'}\mathbf{F}_t^{(k)}}^2 + \widehat{\bs{\lambda}}_i^{(k+1)'}\mathbf{P}_{t\giv t-1}^{(k)}\widehat{\bs{\lambda}}_i^{(k+1)}}
\end{align*}
where $\bs{\lambda}_i$ is the $i$-th row of the matrix of stacked loadings $\bs{\Lambda} = (\bs{\lambda}_1,\ldots,\bs{\lambda}_{n})'$.
\item \textbf{Parameters of the law of motion of the common factors}:
\begingroup
\allowdisplaybreaks
\begin{align*}
\hat{\mathbf{A}}^{(s),(k+1)}\ &=\ \parT{\sum_{t=2}^T \mathbf{f}_{t\giv T}^{(s),(k)}\mathbf{f}_{t-1\giv T}^{(s),(k)^{\prime}} + \mathbf{P}_{t,t-1\giv T}^{\mathbf{f}^{(s)},(k)}}\parT{\sum_{t=2}^T \mathbf{f}_{t-1\giv T}^{(s),(k)}\mathbf{f}_{t-1\giv T}^{(s),(k)^{\prime}} + \mathbf{P}_{t-1\giv T}^{\mathbf{f}^{(s)},(k)}}^{-1}\\[5pt]
\hat{\mathbf{A}}^{(c,s),(k+1)}\ &=\ \parT{\sum_{t=2}^T \mathbf{f}_{t\giv T}^{(c,s),(k)}\mathbf{f}_{t-1\giv T}^{(c,s),(k)^{\prime}} + \mathbf{P}_{t,t-1\giv T}^{\mathbf{f}^{(c,s)},(k)}}\parT{\sum_{t=2}^T \mathbf{f}_{t-1\giv T}^{(c,s),(k)}\mathbf{f}_{t-1\giv T}^{(c,s),(k)^{\prime}} + \mathbf{P}_{t-1\giv T}^{\mathbf{f}^{(c,s)},(k)}}^{-1}\\[5pt]
\widehat{\boldsymbol{\Sigma}}_u^{(s),(k+1)}\ &=\ \frac{1}{T}\left[\sum_{t=2}^T\left(\mathbf{f}_{t \mid T}^{(s),(k)} \mathbf{f}_{t \mid T}^{(s),(k)^{\prime}}+\mathbf{P}_{t \mid T}^{\mathbf{f}^{(s)},(k)}\right)-\hat{\mathbf{A}} \sum_{t=2}^T\left(\mathbf{f}_{t \mid T}^{(s),(k)} \mathbf{f}_{t-1 \mid T}^{(s),(k)^{\prime}}+\mathbf{P}_{t, t-1 \mid T}^{\mathbf{f}^{(s)},(k)}\right)\right]\\[5pt]
\widehat{\boldsymbol{\Sigma}}_u^{(c,s),(k+1)}\ &=\ \frac{1}{T}\left[\sum_{t=2}^T\left(\mathbf{f}_{t \mid T}^{(c,s),(k)} \mathbf{f}_{t \mid T}^{(c,s),(k)^{\prime}}+\mathbf{P}_{t \mid T}^{\mathbf{f}^{(c,s)},(k)}\right)-\hat{\mathbf{A}} \sum_{t=2}^T\left(\mathbf{f}_{t \mid T}^{(c,s),(k)} \mathbf{f}_{t-1 \mid T}^{(c,s),(k)^{\prime}}+\mathbf{P}_{t, t-1 \mid T}^{\mathbf{f}^{(c,s)},(k)}\right)\right]
\end{align*}
\endgroup
\end{enumerate}

\subsubsection{Convergence}
\label{sbsbsec::EMconv}

At each iteration $k$ of the algorithm, an estimate of the log-likelihood conditional on the estimated parameters, $\ell(\mathbf{X},\widehat{\bs{\theta}}^{(k)})$, is obtained from the Kalman Filter. To assess the convergence of the EM algorithm, we then employ the likelihood-based criterion introduced by \citet{doz2012quasi}, based on the (scaled) difference between the log-likelihood across subsequent iterations of the algorithm:\\
\begin{equation*}
\Delta \ell^{(k+1)}=\frac{\left|\ell\left(\mathbf{X} ; \widehat{\boldsymbol{\theta}}^{\left(k+1\right)}\right)-\ell\left(\mathbf{X}; \widehat{\boldsymbol{\theta}}^{\left(k\right)}\right)\right|}{\frac{1}{2}\left|\ell\left(\boldsymbol{X}; \widehat{\boldsymbol{\theta}}^{\left(k+1\right)}\right)+\ell\left(\mathbf{X}; \widehat{\boldsymbol{\theta}}^{\left(k\right)}\right)\right|}
\end{equation*}\\
The algorithm is stopped when $\Delta \ell^{(k+1)} < \varepsilon$, for a predetermined threshold $\varepsilon$, which is set to $10^{-3}$.

\subsubsection{Confidence Intervals}
\label{sbsbsec::confint}
In this paper, we follow the procedure outlined in \citet{barigozzi2022estimation} to compute confidence intervals for the factors estimated via the EM algorithm, properly modified so to account for the block structure within the model. Under the assumption of cross-sectionally uncorrelated idiosyncratic components the variance-covariance matrix of the common and country-specific factors can be estimated as:
\begin{align*}
\tilde{\bs{\Gamma}}^{(s)}\ &=\ \parT{\frac{1}{N^{(s)}} \sum_{i\in N^{(s)}}\widehat{\bs{\gamma}}_{i}'(\widehat{\sigma}_{\xi_{i}}^{2})^{-1}\widehat{\bs{\gamma}}_{i}}^{-1}\\[5pt]
\tilde{\bs{\Gamma}}^{(c,s)}\ &=\ \parT{\frac{1}{N^{(c,s)}} \sum_{i\in N^{(c,s)}}\widehat{\bs{\phi}}_{i}'(\widehat{\sigma}_{\xi_{i}}^{2})^{-1}\widehat{\bs{\phi}}_{i}}^{-1} 
\end{align*}\\
where $\widehat{\sigma}_{\xi_{i}}^{2}$ is the estimated variance of the idiosyncratic component of the $i$-th variable.\footnote{We also try implementing the adaptive thresholding procedure as outlined in \citet{cai2011adaptive}, finding that only $\approx 3\%$ of outer-diagonal entries are actually shrinked to zero, yielding nearly identical results. Indeed, as shown also in Appendix \ref{app:modspec}, the block-structure is able to account for all relevant comovements in the panel, including local dynamics, which leave almost no relevant information in the idiosyncratic component} 

Although consistent, these estimates are likely to misrepresent the true uncertainty on the underlying factors. Indeed, as pointed out by \citet{maldonado2021accurate}, the estimated variance-covariance matrix of the factors depends on the loadings, which are themselves estimated. Hence, we need to account for the additional uncertainty which stems from the estimation of the loadings. This is particularly important in our setting, where we stress the factors at their 95\% significance level. Failing to account for loadings uncertainty could make our results either upward or downward biased, resulting in an over- or under-estimation of the true level of stress for the factors.
 
In the paper, we follow the subsampling procedure proposed by \citet{maldonado2021accurate}, properly modified to account for the block structure of our ML-DFM.   Specifically, for each country-sector subgroup, we draw randomly, without replacement, a $T\times N_{c,s}^*$ sub-matrix of data, say $X_{t}^{\scaleto{(c,s)}{6pt}^*}$. The size of the sub-matrix is selected in a data-driven way, such that $N_{c,s}^* = p^{(c,s)}\cdot N^{(c,s)}$, where $p^{(c,s)} =  1 - \frac{235}{(N^{(c,s)} + 25)^{2}} - 0.2\frac{\sqrt{N^{(c,s)}}}{T} - \frac{r^{c,s}}{N^{(c,s)}T}$. Given the sub-blocks for each country-sector pair, we then aggregate them to obtain the corresponding sub-blocks at the sectoral level, used to extract the sector-specific factors common to all countries. The reason for this \enquote{bottom-down} procedure is to avoid over- or under-representation of countries or sectors, which would be highly likely if we resampled the data being agnostic of the block-structure, especially since macroeconomic variables are sensibly more than financial ones.

Once the sub-matrix of data is selected, we extract the factors via PC, used to initialize the EM algorithm, which is run until convergence. This procedure is repeated $B$ times and, at the end of each iteration, a new estimated of the factors, denoted as $\widehat{\mathbf{F}}_t^{(b)}$ for $b=1,\ldots,B$, is stored. In the paper we set $B=199$ repetitions, as results are nearly identical using more replications, only being more computational demanding. Given the factors estimated at each iteration $b$, an estimate of the variance-covariance matrix of the factors which accounts for the uncertainty in the estimation of the loadings, can be finally computed. Specifically we define as $\widehat{\bs{\Gamma}}^{(s)}$ the estimated variance-covariance matrix for the sector-specific factors common to all countries, and with $\widehat{\bs{\Gamma}}^{(c,s)}$ the estimated variance-covariance matrix for the country-sector specific factors. Then:\\
\begin{align*}
\widehat{\bs{\Gamma}}^{(s)}\ =\ \tilde{\bs{\Gamma}}^{(s)} + \frac{N^{(s)^*}}{N^{(s)}BT}\sum_{b=1}^{B}\sum_{t=1}^{T}\parT{\widehat{\mathbf{f}}_t^{(s),(b)} - \widehat{\mathbf{f}}_t^{(s)}}\parT{\widehat{\mathbf{f}}_t^{(s),(b)} - \widehat{\mathbf{f}}_t^{(s)}}'\\[5pt]
\widehat{\bs{\Gamma}}^{(c,s)}\ =\ \tilde{\bs{\Gamma}}^{(c,s)} + \frac{N^{(c,s)^*}}{N^{(c,s)}BT}\sum_{b=1}^{B}\sum_{t=1}^{T}\parT{\widehat{\mathbf{f}}_t^{(c,s),(b)} - \widehat{\mathbf{f}}_t^{(s)}}\parT{\widehat{\mathbf{f}}_t^{(c,s),(b)} - \widehat{\mathbf{f}}_t^{(c,s)}}'
\end{align*}

\newpage 
\section{Model Specification}
\label{app:modspec}
\begin{figure}[ht!]
\caption{Idiosyncratic cross-correlation: $q=1$ for each block} 
\label{fig::idcorr_q1}
\centering \footnotesize \sc
\setlength{\tabcolsep}{.01\textwidth}    
\begin{tabular}{c}
\includegraphics[width = 0.9\textwidth]{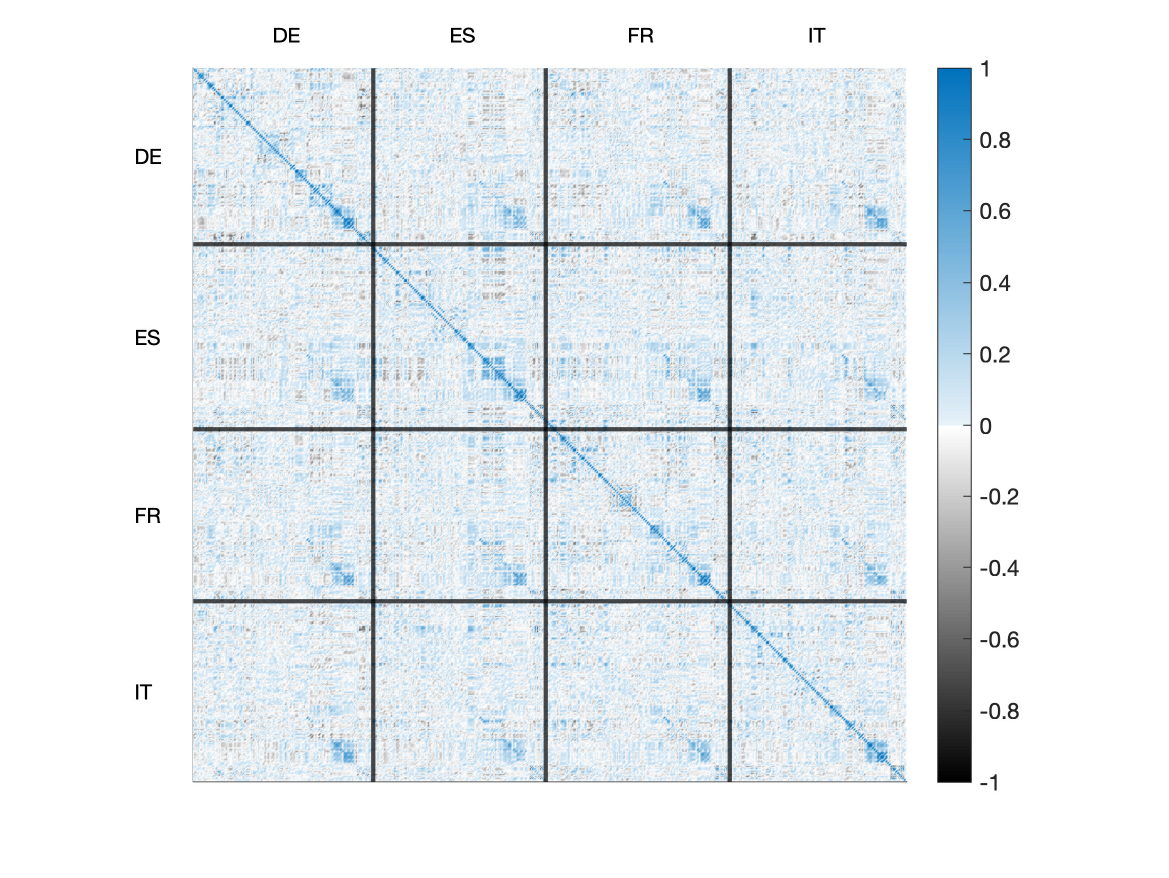} 
\end{tabular}    
\end{figure}

The selection of the number of factors of the ML-DFM is carried out by analysing the structure of the sample covariance matrix of the idiosyncratic residuals obtained when estimating Model \eqref{eq::mlDFM} via the EM algorithm with the most parsimonious specification, i.e. with just one global and one country-specific factor for each country, i.e. $\widehat{\bs{\Xi}} = (\widehat{\bs{\xi}}_{1}',\ldots,\widehat{\bs{\xi}}_{N}')'$. Figure \ref{fig::idcorr_q1} is a heat-map of the corresponding pairwise cross-correlations. As we can see, although the most parsimonious model, with one country-specific factor, is able to capture the bulk of within-sector dynamics, there are still some small blocks of cross-correlated idiosyncratic components. This may be particularly relevant as the blocks are not randomly located, but they are concentrated around the same group of series for all countries (bottom-right corner). It turns out that these are mostly price series. Hence, it seems that one factor for macroeconomic variables is not able to account for some of the comovements across price series, either within or across countries (or both).

To account for the presence of cross-correlation across price series, we consider estimating an additional global macroeconomic factor and, alternatively, an country-specific additional macroeconomic factor for each country. If the price dynamics are mostly idiosyncratic within countries, then an additional macroeconomic factor for each country will incorporate this information.

\begin{figure}[ht!]
\caption{Idiosyncratic cross-correlation: alternative model specifications} 
\label{fig::idcorr_qMC}
\centering \footnotesize \sc
\setlength{\tabcolsep}{.01\textwidth}    
\begin{tabular}{cc}
\scriptsize Common macroeconomic factor & \scriptsize Country-specific macroeconomic factors\\
\includegraphics[trim={2cm 0cm 2cm 0cm},clip,width = 0.5\textwidth]{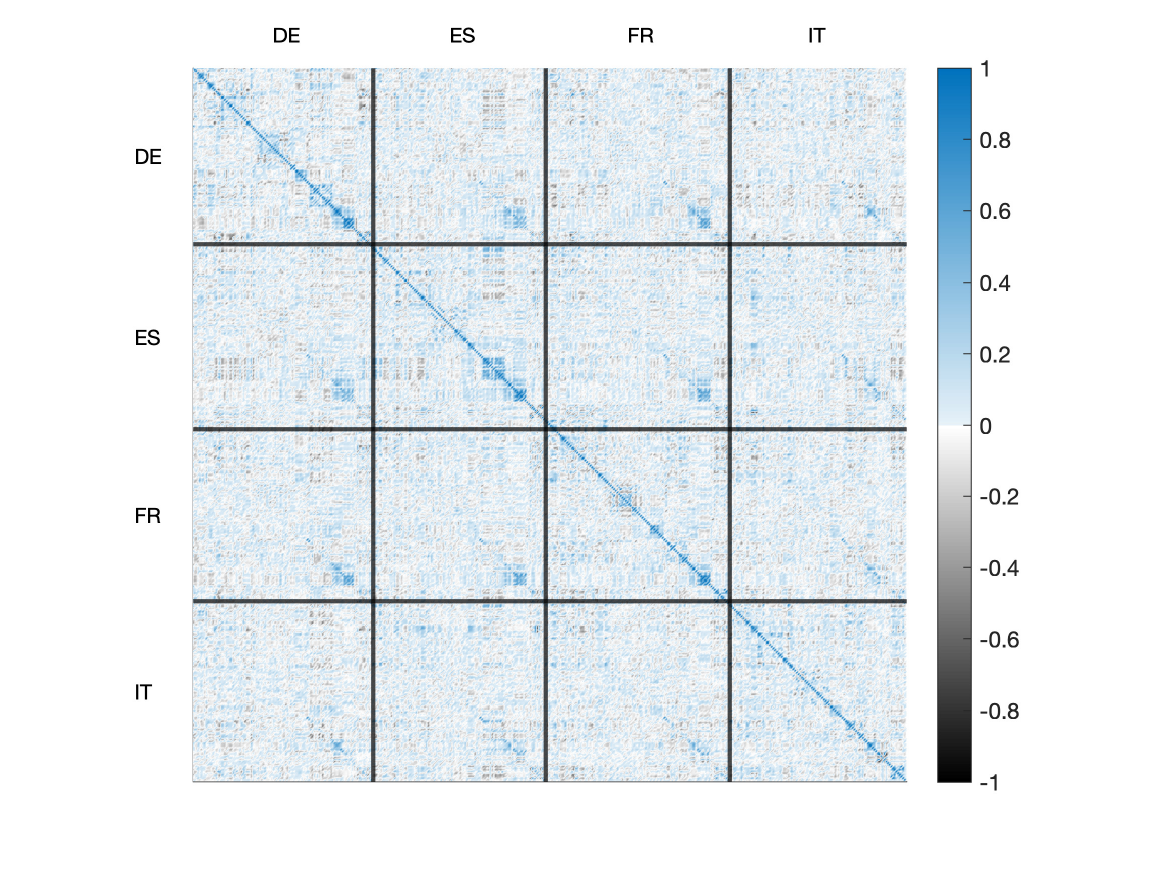} & \includegraphics[trim={2cm 0cm 2cm 0cm},clip,width = 0.5\textwidth]{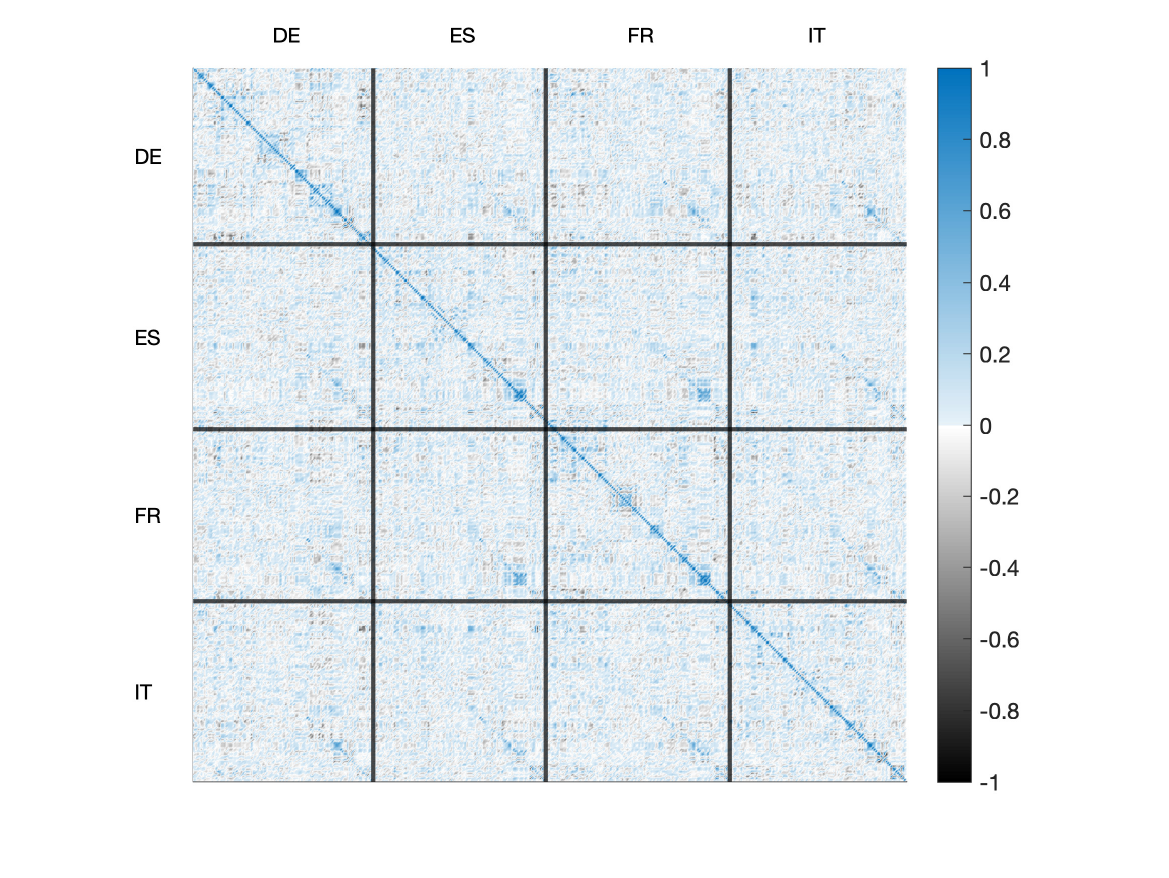} 
\end{tabular}    
\end{figure}

Figure \ref{fig::idcorr_qMC} plots the matrix of cross-correlations across estimated idiosyncratic components in both the cases outlined above. From the graph, it emerges that one additional factor for each country seems to be the most appropriate solution to account for the presence of local dynamics within price series.

\end{document}